\documentclass[11pt]{article}
\usepackage{axodraw}
\usepackage{epsfig}
\usepackage{amsfonts}
\usepackage{amsmath}
\usepackage{bm,bbm}
\allowdisplaybreaks[1]

\input pix.sty

\newcommand{\VIRTUAL}{($2\!\to\! 1$)}
\newcommand{\REAL}{($2\!\to\! 2$)}

\newcommand{\ko}{k_0}
\newcommand{\km}{k_-}
\newcommand{\kp}{k_+}
\newcommand{\aL}{a^{ }_\rmii{L}}
\newcommand{\aR}{a^{ }_\rmii{R}}
\renewcommand{\eq}{eq.~}
\renewcommand{\eqs}{eqs.~}
\renewcommand{\se}{sec.~}
\renewcommand{\ses}{secs.~}
\renewcommand{\fig}{fig.~}
\renewcommand{\figs}{figs.~}

\newcommand{\Nc}{N_{\rm c}}

\newcommand{\gammaE}{\gamma_\rmii{E}}

\newcommand{\rmO}{{\mathcal{O}}}
\newcommand{\bmu}{\bar\mu}

\def\lsi{\raise0.3ex\hbox{$<$\kern-0.75em\raise-1.1ex\hbox{$\sim$}}}
\def\gsi{\raise0.3ex\hbox{$>$\kern-0.75em\raise-1.1ex\hbox{$\sim$}}}
\newcommand{\lsim}{\mathop{\lsi}}
\newcommand{\gsim}{\mathop{\gsi}}

\newcommand{\nF}{n_\rmii{F}}
\newcommand{\nB}{n_\rmii{B}}
 \renewcommand{\nF}[1]{n_\rmii{F{#1}}}
 \renewcommand{\nB}[1]{n_\rmii{B{#1}}}

\newcommand{\rmii}[1]{{\mbox{\tiny\rm{#1}}}}

\newcommand{\im}{\mathop{\mbox{Im}}}

\newcommand{\Tint}[1]{{\hbox{$\sum$}\!\!\!\!\!\!\!\int\,}_{\!\!\!\!\raise-0.9ex\hbox{$\scriptstyle{#1}$}}}
\newcommand{\Tinti}[1]{{{\Sigma}\!\!\!\!\raise0.3ex\hbox{$\int$}_\rmii{${#1}$}}}

\newcommand{\bi}{\begin{itemize}}
\newcommand{\ei}{\end{itemize}}
\newcommand{\hide}[1]{ }
\newcommand{\bsl}[1]{\,\slash\!\!\!\!{#1}\,}

\def\TAsc(#1,#2)(#3,#4,#5)%
{\SetWidth{2.0}\CArc(#1,#2)(#3,#4,#5)\SetWidth{1.0}}
\def\Lwidth{3}

\def\TAgl(#1,#2)(#3,#4,#5){\SetWidth{2.0}\PhotonArc(#1,#2)(#3,#4,#5){\Lwidth}%
{6.283 #3 mul 360 div #4 #5 sub #4 #5 sub mul sqrt mul Tdensity mul}%
\SetWidth{1.0}}
\def\TLgl(#1,#2)(#3,#4){\SetWidth{2.0}\Photon(#1,#2)(#3,#4){\Lwidth}
{#1 #3 sub #1 #3 sub mul #2 #4 sub #2 #4 sub mul add sqrt Tdensity mul}%
\SetWidth{1.0}}
\renewcommand{\picc}[1]{\;\parbox[c]{60pt}{\begin{picture}(60,30)(0,0)
\SetWidth{1.0}\SetScale{1.3} #1 \end{picture}}\; }
\def\Lwidth{1.3}

\newcommand{\picg}[1]{\;\parbox[c]{62pt}{\begin{picture}(160,80)(0,0)
\SetWidth{1.0}\SetScale{0.5} #1 \end{picture}}\;}

 \def\GraphLxxii{\picg{
  \SetWidth{1.5}
  \DashLine(0,60)(40,40){4} 
  \Line(0,20)(40,40) 
  \Line(40,42)(80,42)
  \Line(40,38)(80,38)
  \Photon(31,44.5)(31,64.5){2.3}{3}
  \Photon(31,15.5)(31,35.5){2.3}{3}
  \Line(31,64.5)(21,69.5)
  \Line(31,64.5)(41,69.5)
  \Line(31,15.5)(21,10.5)
  \Line(31,15.5)(41,10.5)
  \GCirc(18,29){3.5}{0}
  \GCirc(18,51){3.5}{0}
  \GCirc(40,40){3.5}{0}
  \GCirc(31,54.5){3.5}{0}
  \GCirc(31,25.5){3.5}{0}
 }}
 \def\GraphLxxiii{\picg{
  \SetWidth{1.5}
  \DashLine(0,60)(40,40){4} 
  \Line(0,20)(40,40) 
  \Line(40,42)(80,42)
  \Line(40,38)(80,38)
  \GCirc(18,29){3.5}{0}
  \GCirc(18,51){3.5}{0}
  \GCirc(40,40){3.5}{0}
 }}
 \def\GraphLxxiv{\picg{
  \SetWidth{1.5}
  \Lqu(0,20)(36,38) 
  \DashLine(0,60)(36,42){4} 
  \Line(36,42)(80,42)
  \Line(36,38)(80,38)
 }}
 \def\GraphLxxv{\picg{
  \SetWidth{1.5}
  \Line(80,62)(40,42) 
  \Line(80,58)(40,38) 
  \Line(80,20)(40,40) 
  \DashLine(40,40)(0,40){4}
  \Photon(30,40)(30,60){2.3}{3}
  \Photon(49,15.5)(49,35.5){2.3}{3}
  \Line(30,60)(40,65)
  \Line(30,60)(20,65)
  \Line(49,15.5)(59,10.5)
  \Line(49,15.5)(39,10.5)
  \GCirc(62,29){3.5}{0}
  \GCirc(18,40){3.5}{0}
  \GCirc(40,40){3.5}{0}
  \GCirc(30,50){3.5}{0}
  \GCirc(49,25.5){3.5}{0}
 }}
 \def\GraphLxxvi{\picg{
  \SetWidth{1.5}
  \Line(80,62)(40,42) 
  \Line(80,58)(40,38) 
  \Line(80,20)(40,40) 
  \DashLine(40,40)(0,40){4}
  \GCirc(62,29){3.5}{0}
  \GCirc(18,40){3.5}{0}
  \GCirc(40,40){3.5}{0}
 }}
\def\procA{\picb{%
 \Lqu(0,0)(24,14)%
 \Lsc(0,30)(24,16)%
 \Line(24,14)(40,14)%
 \Line(24,16)(40,16)%
}}
\def\procB{\picb{%
 \Lqu(0,0)(15.4,9)%
 \Line(0,0)(24,14)%
 \Lgl(12,7)(12,23)%
 \Lsc(0,30)(24,16)%
 \Line(24,14)(40,14)%
 \Line(24,16)(40,16)%
}}
\def\procC{\picb{%
 \Lqu(0,0)(24,14)%
 \Lsc(0,30)(24,16)%
 \Agl(12,23)(5,-40,140)%
 \Line(24,14)(40,14)%
 \Line(24,16)(40,16)%
}}
\def\procCa{\picb{%
 \Lqu(0,0)(24,14)%
 \Lsc(0,30)(24,16)%
 \Agl(16,26.5)(5,-135,225)%
 \Line(24,14)(40,14)%
 \Line(24,16)(40,16)%
}}
\def\procCb{\picb{%
 \Lqu(0,0)(24,14)%
 \Lsc(0,30)(6.8,26)%
 \Lsc(18,19.5)(24,16)%
 \Aqu(12,23)(5,-45,135)%
 \Aqu(12,23)(5,135,315)%
 \Line(24,14)(40,14)%
 \Line(24,16)(40,16)%
}}
\def\procCc{\picb{%
 \Lqu(0,0)(24,14)%
 \Lsc(0,30)(24,16)%
 \Asc(16,26.5)(5,-135,225)%
 \Line(24,14)(40,14)%
 \Line(24,16)(40,16)%
}}
\def\procD{\picb{%
 \Lqu(0,0)(24,14)%
 \Lsc(0,30)(24,16)%
 \Agl(12,7)(5,-145,35)%
 \Line(24,14)(40,14)%
 \Line(24,16)(40,16)%
}}
\def\procE{\picb{%
 \Lqu(0,0)(24,14)%
 \Lsc(0,30)(24,16)%
 \Lgl(0,13)(15.4,21)%
 \Line(24,14)(40,14)%
 \Line(24,16)(40,16)%
}}
\def\procEa{\picb{%
 \Laqu(0,13)(15.4,21)%
 \Lqu(0,30)(15.4,21)%
 \Lsc(15.4,21)(24,16)%
 \Lqu(0,0)(24,14)%
 \Line(24,14)(40,14)%
 \Line(24,16)(40,16)%
}}
\def\procF{\picb{%
 \Lqu(0,0)(15.4,9)%
 \Line(0,0)(24,14)%
 \Lgl(0,18)(15.4,9)%
 \Lsc(0,30)(24,16)%
 \Line(24,14)(40,14)%
 \Line(24,16)(40,16)%
}}
\def\procG{\picb{%
 \Lqu(0,30)(15,15)%
 \Lgl(0,0)(15,15)%
 \Lqu(15,15)(29,15)%
 \Lsc(31,14)(45,0)%
 \Line(29,15)(44,30)%
 \Line(31,14)(45,28)%
}}
\def\procGa{\picb{%
 \Lsc(0,30)(15,15)%
 \Lgl(0,0)(15,15)%
 \Lsc(15,15)(29,15)%
 \Laqu(31,14)(45,0)%
 \Line(29,15)(44,30)%
 \Line(31,14)(45,28)%
}}
\def\procH{\pic{%
 \Lqu(0,30)(15,26)%
 \Lgl(0,0)(15,4)%
 \Lsc(15,4)(15,24)%
 \Lsc(15,4)(30,0)%
 \Line(15,24)(30,28)%
 \Line(14,26)(30,30)%
}}
\def\procHa{\pic{%
 \Lsc(0,30)(15,26)%
 \Lgl(0,0)(15,4)%
 \Lqu(15,4)(15,24)%
 \Laqu(15,4)(30,0)%
 \Line(15,24)(30,28)%
 \Line(14,26)(30,30)%
}}
\def\procI{\pic{%
 \Lqu(0,0)(15,4)%
 \Lsc(0,30)(15,26)%
 \Lqu(15,4)(15,24)%
 \Lgl(15,4)(30,0)%
 \Line(15,24)(30,28)%
 \Line(14,26)(30,30)%
}}
\def\procJ{\pic{%
 \Lqu(0,30)(15,26)%
 \Lsc(0,0)(15,4)%
 \Lsc(15,4)(15,24)%
 \Lgl(15,4)(30,0)%
 \Line(15,24)(30,28)%
 \Line(14,26)(30,30)%
}}
\def\procKa{\picb{%
 \Lqu(0,30)(15,15)%
 \Laqu(0,0)(15,15)%
 \Lsc(15,15)(29,15)%
 \Laqu(31,14)(45,0)%
 \Line(29,15)(44,30)%
 \Line(31,14)(45,28)%
}}
\def\procKb{\pic{%
 \Lqu(0,30)(15,26)%
 \Lqu(0,0)(15,4)%
 \Lsc(15,4)(15,24)%
 \Lqu(15,4)(30,0)%
 \Line(15,24)(30,28)%
 \Line(14,26)(30,30)%
}}
\def\procKc{\pic{%
 \Lqu(0,30)(15,26)%
 \Laqu(0,0)(15,4)%
 \Lsc(15,4)(15,24)%
 \Laqu(15,4)(30,0)%
 \Line(15,24)(30,28)%
 \Line(14,26)(30,30)%
}}
\makeatletter \@addtoreset{equation}{section} \makeatother
\renewcommand{\theequation}{\arabic{section}.\arabic{equation}}
\makeatletter
\renewcommand\section{\@startsection {section}{1}{\z@}%
                                   {-5.5ex \@plus -1ex \@minus -.2ex}% bfr-
                                   {2.3ex \@plus.2ex}%
                                   {\normalfont\large\bfseries}}
\renewcommand\subsection{\@startsection{subsection}{2}{\z@}%
                                     {-3.25ex\@plus -1ex \@minus -.2ex}%
                                     {1.5ex \@plus .2ex}%
                                     {\normalfont\normalsize\bfseries}}
\renewcommand\thesection {\@arabic\c@section}
\renewcommand\thesubsection   {\thesection.\@arabic\c@subsection}
\renewcommand{\@seccntformat}[1]{%
\csname the#1\endcsname.\hspace{1.0em}}
\makeatother
\begin{document}

\flushbottom

\begin{titlepage}

\begin{flushright}
% Notes M.L. \\ 
% arXiv:1411.1765\\ 
\vspace*{1cm}
\end{flushright}
\begin{centering}
\vfill

{\Large{\bf
 Right-handed neutrino production rate at $T > 160$~GeV
}} 

\vspace{0.8cm}

I.~Ghisoiu$^{\rm a}$ and M.~Laine$^{\rm b}$ 

\vspace{0.8cm}

$^\rmi{a}$%
{\em
Department of Physics and Helsinki Institute of Physics, \\
University of Helsinki, P.O.Box 64,  
FI-00014 Helsinki, Finland\\}

\vspace{0.3cm}

$^\rmi{b}$%
{\em
Institute for Theoretical Physics, 
Albert Einstein Center, University of Bern, \\ 
Sidlerstrasse 5, CH-3012 Bern, Switzerland\\}

% \vspace*{0.3cm}

\vspace*{0.8cm}

\mbox{\bf Abstract}
 
\end{centering}

\vspace*{0.3cm}
 
\noindent
The production rate of right-handed neutrinos from a Standard Model plasma at
a temperature above a hundred GeV has previously been evaluated up to NLO in
Standard Model couplings ($g\sim 2/3$) in relativistic ($M \sim \pi T $) and 
non-relativistic regimes ($M \gg \pi T$), and up to LO in an ultrarelativistic
regime ($M \, \lsim \, gT$). The last result necessitates an all-orders 
resummation of the loop expansion, accounting for multiple soft scatterings 
of the nearly light-like particles participating in $1\leftrightarrow 2$
reactions. In this paper we suggest how the regimes can be interpolated 
into a result applicable for any right-handed neutrino mass and at all 
temperatures above 160 GeV. The results can also be used for determining 
the lepton number washout rate in models containing right-handed neutrinos. 
Numerical results are given in a tabulated form permitting for their 
incorporation into leptogenesis codes. We note that due to effects from 
soft Higgs bosons there is a narrow intermediate regime around
$M \, \sim \, g^{1/2}T$
in which our interpolation is phenomenological and 
a more precise study would be welcome. 

\vfill

%% %\noindent
%% %PACS numbers: 
%% %11.10.Wx, %        Finite temperature field theory
%% { %11.15.Ha, %        Lattice gauge theory } 
%% %12.38.Bx, %        Perturbative calculations in QCD
%% %12.38.Mh, %        Quark--gluon plasma
%% %14.40.Nd, %        Bottom mesons
%% %\\
%% %Keywords: Thermal Field Theory, Neutrino Physics, Resummation
 
\vspace*{1cm}
  
\noindent
November 2014

\vfill

\end{titlepage}

%%%%%%%%%%%%%%%%%%%%%%%%%%%%% SECTION %%%%%%%%%%%%%%%%%%%%%%%%%%%%%%%%%%%%
%
\section{Introduction}

The Standard Model completed by several generations of right-handed
neutrinos represents a minimal renormalizable framework which is 
able to describe
all available data from terrestrial experiments. 
It appears well 
motivated to explore the cosmological significance of this 
framework~\cite{yanagida}--\cite{numsm_rev}. 
The present paper aims 
to contribute to such an endeavour, by studying the behaviour of 
right-handed neutrinos of any mass $M$ 
(1~GeV $\lsim \, M \, \lsim\, 10^{15}$~GeV). 
We concentrate
on temperatures above about 160~GeV, so that the Higgs mechanism 
is not operative and the vacuum masses of particles such as
the top quark or $W^\pm,Z^0$ bosons can be neglected. The 
right-handed neutrinos interact with the Standard Model degrees of 
freedom through Yukawa interactions, which are assumed to be weaker
than typical Standard Model interactions (gauge interactions, 
or Yukawa interactions associated with the top quark).
We treat the neutrino Yukawa interactions at leading order, whereas 
for Standard Model interactions the goal is to 
explore the magnitude of higher-order corrections as well. 

It is well known that relativistic thermal field theories
suffer from a breakdown of the conventional
loop expansion. One reason is that multiple interactions in the plasma
generate thermal masses for different excitations, thereby 
forming a system of ``quasiparticles'' whose kinematics may differ
substantially from the vacuum case. Another aspect of the problem is that
when a highly energetic particle passes through a plasma, very many 
interactions take place within the time scale needed for the initial particle
to decay or coalesce into another particle. It is a particular goal 
of the present paper to derive results which in one limit extrapolate
to the validity range of the conventional loop expansion, and in 
another to a regime where the mentioned thermal
effects need to be systematically ``resummed''. 

More precisely, one observable we consider is the production rate
of right-handed neutrinos from an initial state in which the Standard Model 
particles are in equilibrium at a temperature $T$,
whereas the right-handed neutrinos
appear with an abundance much smaller than the equilibrium one. 
There have been several recent studies of this production 
rate. In the so-called non-relativistic 
regime~\cite{salvio,nonrel,tum}, the conventional loop expansion
does apply, with thermal corrections appearing only
through small power corrections $\sim \rmO(T^2/M^2)$~\cite{sch}. 
Alas, the non-relativistic
expansion shows convergence only at very 
low temperatures $T \lsim M/15$~\cite{relat}, 
by which time most of the physics of interest to, say, leptogenesis, 
has already played its role~\cite{bw}. A broader ``relativistic'' 
regime $ T \lsim M/3$ can also be addressed up to next-to-leading order
(NLO), even if only in numerical form~\cite{relat}.
In the relativistic regime the conventional 
loop expansion is still valid at NLO. Increasing the temperature
to $T \sim M/g^\rmii{1/2}$, where $g$ denotes a generic 
Standard Model coupling, the loop expansion breaks down for the 
first time. This initial breakdown can be cured by resumming
a subset of higher order diagrams into a thermal mass for the 
Higgs field~\cite{relat}. Proceeding to $T \gsim M/g$, a further 
reorganization is needed. Then an iterated resummation, 
including Hard Thermal Loop (HTL) resummation for propagators and
vertices, and Landau-Pomeranchuk-Migdal (LPM) resummation accounting
for multiple soft scatterings, needs to be 
implemented in order to obtain correct
leading-order (LO) results~\cite{bb0,bb1}. 
As is usual with effective descriptions, the LPM result needs
to be systematically combined with other (non-resummed) 
processes contributing at the same order~\cite{bb2}.  
Borrowing effective field theory language, we refer to the latter step
as a ``matching computation''. 
In its current implementation the matching computation is  
{\em only} valid for the ultrarelativistic 
temperatures $T \gsim M/g$, 
because it was carried out by setting $M/T=0$.
It is also discomforting that another 
(phenomenological) implementation of the
matching computation led to a much larger production rate~\cite{bg2}.

The objective
of the present paper is to suggest a smooth interpolation between 
the relativistic regime $T \lsim M/3$ and the 
LPM-resummed ultrarelativistic regime $T \gsim M/g$. 
We present numerical results in a tabulated form 
which hopefully permits for their practical incorporation 
into leptogenesis computations. 

The plan of this paper is the following. 
After defining the observables considered in \ref{se:defs}, 
existing NLO computations and LPM resummations are reviewed in 
\ses\ref{se:nlo} and \ref{se:lpm}, respectively. 
In \se\ref{se:subtraction} the NLO result is dissected into 
a contribution also appearing as a part of LPM resummation, 
and a remainder which needs to be added to the LPM result. 
An important subtlety, stemming from the fact that 
the NLO computation reviewed in \se\ref{se:nlo}
was carried out with a vanishing (thermal) Higgs mass, 
is addressed in \se\ref{se:X}. All ingredients
are put together in \se\ref{se:interpolate}, 
leading to a single result interpolating between different regimes
(however the interpolant's 
parametric accuracy varies from regime to regime). 
Some conclusions and an outlook are offered in \se\ref{se:concl}.
Four appendices contain details concerning the NLO result,
HTL resummation, cancellation of infrared divergences, 
and choice of parameters. 

%%%%%%%%%%%%%%%%%%%%%%%%%%%%% SECTION %%%%%%%%%%%%%%%%%%%%%%%%%%%%%%%%%%%%
%
\section{Basic definitions}
\la{se:defs}

Denoting by $\mathcal{K} = (\ko,\vec{k})$ the four-momentum of on-shell
right-handed neutrinos of mass $M$, so that $\ko = \sqrt{k^2 + M^2}$
with $k \equiv |\vec{k}|$, and by $h^{ }_{\nu i}(\bmu)$ a renormalized
$\msbar$ neutrino 
Yukawa coupling attaching right-handed neutrinos
to the left-handed lepton generation $i\in\{1,2,3\}$, 
the production rate of out-of-equilibrium right-handed neutrinos
% (per a space-time volume element ${\rm d}^4\mathcal{X}$) 
from an equilibrium plasma at rest can be expressed as
\be
 \frac{{\rm d} N^{ }_+(\mathcal{K})}{{\rm d}^4\mathcal{X} {\rm d}^3\vec{k}}
 \, = \,
 \frac{2\nF{}(\ko)}{(2\pi)^3 \ko} \, |h_\nu|^2 
 \, \im\Pi^{ }_\rmii{R}(\mathcal{K}) + \rmO(|h_\nu|^4)
 \;, \quad
 |h_\nu|^2 \equiv \sum_{i=1}^{3} 
 |h^{ }_{\nu i}(\bmu)|^2  
 \;. \la{rate} 
\ee 
Here $\nF{}$ is the Fermi distribution (similarly, $\nB{}$ denotes
a Bose distribution), and the subscript in $N^{ }_+$ indicates that
both spin states have been summed together. 
The retarded correlator $\Pi^{ }_\rmii{R}$
can be expressed as an analytic continuation of a corresponding
imaginary-time one, 
\be
 \Pi^{ }_\rmii{E}(K) \equiv
 \mathcal{Z}_{\nu}^{ }
 \, \tr \Bigl\{ 
 i \bsl{K} \! \int_0^{1/T} \!\! {\rm d}\tau \int_\vec{x} e^{i K\cdot X}
 \, \Bigl\langle
  (\tilde{\phi}^\dagger  \aL\, \ell)(X) \, 
  (\bar{\ell}\, \aR\, \tilde{\phi} ) (0)
 \Bigr\rangle^{ }_T \Bigr\}
 \;, \la{Pi}
\ee
as
\be
 \Pi^{ }_\rmii{R}(\mathcal{K}) = 
 \left. \Pi^{ }_\rmii{E}(K)
 \right|^{ }_{k_n \to -i [\ko + i 0^+]}
 \;. \la{relation} 
\ee
In these equations, 
$\mathcal{Z}^{ }_{\nu}$ is a renormalization factor related
to the neutrino Yukawa couplings;  
$K \equiv (k_n,\vec{k})$ where $k_n$ denotes a fermionic
Matsubara frequency;
$X \equiv (\tau,\vec{x})$ is a Euclidean space-time coordinate; 
$\tilde{\phi} = i \sigma_2 \phi^*$ is a Higgs doublet;   
$\aL, \aR$ are chiral projectors; 
$\ell$ is a left-handed lepton doublet; 
and $\langle ... \rangle^{ }_T$ denotes an equilibrium expectation value. 
At NLO, the renormalization factor reads 
\be
 \mathcal{Z}^{ }_\nu = 
   1 + \frac{1}{(4\pi)^2\epsilon} 
  \biggl[
    h_t^2 \Nc - \fr34 (g_1^2 + 3 g_2^2)  
  \biggr] + \rmO(g^4)
 \;, \la{Z_nu}
\ee
where the space-time dimension has been expressed
as $D = 4 - 2\epsilon$;
$h_t$ is the renormalized top Yukawa coupling; 
$\Nc \equiv 3$ is the number of colours;  
and $g_1$, $g_2$ are the renormalized hypercharge and weak gauge couplings, 
respectively. 
In addition to these couplings, the Higgs self-coupling
$\lambda$ also appears in our results; how these parameters are fixed
in terms of physical observables 
is explained in appendix~D. The notation $g^2$ refers generically
to the couplings $h_t^2, g_1^2, g_2^2, \lambda$ which are taken
to be parametrically of the same order of magnitude. 

The total production rate of right-handed neutrinos reads
\be
 \frac{\gamma_+ (M)}{|h_\nu|^2} = 
 \int_\vec{k} \frac{2\nF{}(\ko)}{\ko} \, 
% \left. 
  \im\Pi^{ }_\rmii{R}(\mathcal{K}) % \right|_{\mathcal{K}^2 = M^2}
 + \rmO(|h_\nu|^2)
 \;, \la{gamma} 
\ee
where the integration measure is defined as 
$
 \int_\vec{k} \equiv \int \! \frac{{\rm d}^3\vec{k}}{(2\pi)^3}
$.
A closely related quantity determines the lepton number 
dissipation (``washout'') rate
in models containing right-handed neutrinos~\cite{washout}: 
\be
 \mathcal{W}(M) \equiv 
 - \int_\vec{k} \frac{2\nF{}'(\ko)}{\ko} \, 
% \left.
  \im\Pi^{ }_\rmii{R}(\mathcal{K}) % \right|_{\mathcal{K}^2 = M^2}
% + \rmO(|h_\nu|^2) 
 \;. \la{washout}
\ee
As explained in ref.~\cite{washout}, 
this needs to be combined with a ``susceptibility matrix'' 
and a group-theoretic prefactor of $\rmO(|h_\nu|^2)$ 
in order to get the complete result for the washout rate. 
% (the susceptibility matrix has recently been analyzed
% in ref.~\cite{sangel}).

Rather than the differential rates 
$\partial_k \gamma^{ }_+$ and $\partial_k \mathcal{W}$, 
we mostly discuss $\im \Pi^{ }_\rmii{R}$ in the following. The reason 
is that, unlike the integrands in \eqs\nr{gamma}, \nr{washout}, 
$\im \Pi^{ }_\rmii{R}$ is a 
Lorentz-invariant quantity in vacuum, i.e.\ only dependent 
on $M^2 = \mathcal{K}^2$ at $T \ll M$ (rather than 
separately on $M$ and $k$). 
At a finite temperature % $T \gsim M$
this is no longer the case (cf.\ \fig\ref{fig:spectraRESUM}(right)), 
however even then $\im \Pi^{ }_\rmii{R}$
turns out to display only a modest dependence on $k$ for fixed $M$, 
and therefore allows us to present results in a relatively economic fashion
(i.e.\ as a sparse table). 

%%%%%%%%%%%%%%%%%%%%%%%%%%%%% SECTION %%%%%%%%%%%%%%%%%%%%%%%%%%%%%%%%%%%%
%
\section{NLO result in the relativistic regime}
\la{se:nlo}

%%%%%%%%%%%%%%%%%%%%%%%%%%%%%%% FIGURE %%%%%%%%%%%%%%%%%%%%%%%%%%%%%%%%%%%%%%
%
\begin{figure}[t]

%\hspace*{1.5cm}%
%\begin{minipage}[c]{3cm}
\begin{eqnarray*}
&& 
 \hspace*{-1cm}
 \procA
 \hspace*{0.1cm}
 \procB
 \hspace*{0.1cm}
 \procC
 \hspace*{0.1cm}
 \procCa
 \hspace*{0.1cm}
 \procCb
 \hspace*{0.12cm}
 \procCc
 \hspace*{0.12cm}
 \procD
 \hspace*{0.12cm}
 \procE
 \hspace*{0.12cm}
 \procF
 \hspace*{0.12cm}
 \procEa
 \\[3mm] 
&& 
 \hspace*{-1cm}
 \procG
 \hspace*{0.55cm}
 \procGa
 \hspace*{0.55cm}
 \procKa
 \hspace*{0.55cm}
 \procH
 \hspace*{0.55cm}
 \procHa
 \hspace*{0.55cm}
 \procI
 \hspace*{0.55cm}
 \procJ
 \hspace*{0.55cm}
 \procKb
 \hspace*{0.55cm}
 \procKc
\end{eqnarray*}
%\end{minipage}

\caption[a]{\small 
 The processes, up to $O(g^2)$, through which right-handed 
 neutrinos can be generated. 
 Arrowed, dashed, and wiggly lines correspond to 
 Standard Model fermions, scalars, and gauge fields, respectively, whereas
 right-handed neutrinos are denoted by a double line. The closed ``virtual''
 loops include both vacuum and thermal corrections. 
} 
\la{fig:processes}
\end{figure}
%
%%%%%%%%%%%%%%%%%%%%%%%%%%%%%%%%%%%%%%%%%%%%%%%%%%%%%%%%%%%%%%%%%%%%%%%%%%%%%

We start by discussing the production rate in the ``naive'' language
of Feynman diagrams and the loop expansion. The relevant amplitudes
are shown in \fig\ref{fig:processes}. If the right-handed neutrino is 
massive and all other particles are assumed massless, the LO process
is the $2\rightarrow 1$ coalescence depicted up left. The NLO level 
includes virtual corrections to the $2\rightarrow 1$ reaction, as well as 
real $3\rightarrow 1$ and $2\rightarrow 2$ processes. In a massless
theory, the real and virtual NLO processes are IR divergent; 
their sum is finite for any $M > 0$~\cite{gh,relat}. 
All the NLO processes have been evaluated
numerically in ref.~\cite{relat}. 

It was pointed out in ref.~\cite{relat}, however,   
that for $M \sim {g}^\rmii{1/2}T$ 
the loop expansion breaks down, and 
a thermal mass resummation is needed for the Higgs field.
The (``asymptotic'') thermal masses associated with 
the Higgs field ($m_\phi^{ }$) and with  
left-handed leptons ($m_\ell^{ }$) are 
\ba
 && m_\phi^2 =  -\frac{m_H^2}{2} + 
  \Bigl( g_1^2 + 3 g_2^2 + \fr43 h_t^2 \Nc + 8 \lambda  
  \Bigr) \frac{T^2}{16}
 \;, \quad
 m_\ell^2 =  \bigl(
   g_1^2 + 3g_2^2 
 \bigr) \frac{T^2}{16} 
 \;, \la{mH}
\ea
where $m_H$ is the vacuum Higgs mass, and 
corrections of $\rmO(g^2 m_H^2, g^3 T^2)$ have been omitted~\cite{sch2}.  
In ref.~\cite{relat} such a mass resummation was 
implemented not only for the Higgs field, for which a resummation 
is unambiguous, but also for leptons, for which
it amounts to a higher-order effect when $M \sim {g}^\rmii{1/2}T$. 
It turns out that once proceeding to $M \lsim gT$, 
where thermal mass resummation becomes necessary 
even for leptons, the correct procedure {\em differs} from the naive
implementation of ref.~\cite{relat} (the correct procedure for leptons 
is part of the LPM resummation as discussed in \se\ref{se:lpm}). Hence, 
in order to be able to combine the NLO result with the LPM result
in a systematic way, we need to 
re-express the NLO result of ref.~\cite{relat} 
{\em without} a thermal mass resummation for leptons. 

Keeping a thermal mass for the Higgs only, the leading-order result with
a general four-momentum $\mathcal{K}$ 
in the time-like domain 
$M^2 \equiv \mathcal{K}^2 > 0$
can be expressed as
\ba
 { \im\Pi_\rmii{R}^\rmii{LO} } % {|h_\nu|^2}
 & \equiv & 
 \frac{(M^2 - m_\phi^2)T}{8\pi k}
 \ln\left\{
   \frac{\sinh\Bigl[ \frac{\kp + {m_\phi^2} / {(4\kp)} }{2T} \Bigr]
   \cosh\Bigl[ \frac{\kp (1 -  {m_\phi^2} / {M^2} ) }{2T} \Bigr] }
  {\sinh \Bigl[ \frac{\km + {m_\phi^2} / {(4\km)} }{2T} \Bigr]
  \cosh\Bigl[ \frac{\km (1 -  {m_\phi^2} / {M^2} ) }{2T} \Bigr]  }
 \right\} 
 \;.
 \la{tree}
\ea
Here we have defined 
\be
 k^{ }_\pm \equiv \frac{\ko \pm k}{2}
 \;. \la{kpm}
\ee 
The result of \eq\nr{tree} can be evaluated (and is positive) 
both for $M > m_\phi$ and $M < m_\phi$.

Once the NLO expression of ref.~\cite{relat} is written as a sum
of \eq\nr{tree} and a remainder, the final result becomes
\ba
 {\im \Pi^\rmii{NLO}_\rmii{R}} % {|h_\nu(\bmu)|^2}
    & \equiv & 
 {\im \Pi^\rmii{LO}_\rmii{R}} % {|h_\nu(\bmu)|^2}
 %%%
 \nn & + &   
 2 h_t^2 \Nc 
 \biggl\{
   - \rho^\rmii{$T$}_{\widetilde{\mathcal{I}}^{ }_\rmii{f}}
   + \rho^\rmii{$T$}_{\widetilde{\mathcal{I}}^{ }_\rmii{h}}
 %%%
 \nn & & \; - \, 
   \frac{\pi {M}^2}{(4\pi)^4 k}
   \int_{\km}^{\kp} \! {\rm d}p \, 
   \frac{ \nF{}(\ko -p) \nB{}(p) }{\nF{}(\ko)}
   \biggl[
     \ln \frac{(\kp - p)(p - \km)\bmu^2 }{k^2 {M}^2} + \fr{11}{2} 
   \biggr]
 \biggr\}
 %%%
 \nn & + &   
 \frac{g_1^2 + 3 g_2^2}{2}
 \biggl\{
     2 \Bigl[ 
%%% NEW
     \rho^\rmii{$T$}_{{\mathcal{I}}^{ }_\rmii{b}}
   - \rho^\rmii{$T$}_{\widetilde{\mathcal{I}}^{ }_\rmii{b}} 
   + \rho^\rmii{$T$}_{\widehat{\mathcal{I}}^{ }_\rmii{d}} 
   - \rho^\rmii{$T$}_{\overline{\mathcal{I}}^{ }_\rmii{d}} 
%%%
    + \rho^\rmii{$T$}_{{\mathcal{I}}^{ }_\rmii{g}}
    + \rho^\rmii{$T$}_{\widehat{\mathcal{I}}^{ }_\rmii{h'}}
    + \rho^\rmii{$T$}_{{\mathcal{I}}^{ }_\rmii{j}}
     \Bigr] 
    - 4 \Bigl[ 
       \rho^\rmii{$T$}_{{\mathcal{I}}^{ }_\rmii{h}}
    + \rho^\rmii{$T$}_{\widehat{\mathcal{I}}^{ }_\rmii{h}}
     \Bigr] 
 \nn & & \; + \, 
   \frac{3 \pi {M}^2}{(4\pi)^4 k}
   \int_{\km}^{\kp} \! {\rm d}p \, 
   \frac{ \nF{}(\ko -p) \nB{}(p) }{\nF{}(\ko)}
   \biggl[
     \ln \frac{(\kp - p)(p - \km)\bmu^2 }{k^2 {M}^2} + \fr{41}{6} 
   \biggr]
 \biggr\}
 \;. \hspace*{10mm} \la{nlo}
\ea
The objects $\rho^\rmii{$T$}_{\mathcal{I}^{ }_\rmii{x}}$ 
are ``master'' spectral functions, depending on $\ko$, $k$
and $T$ and evaluated numerically in refs.~\cite{master,relat}. 
Eq.~\nr{nlo} 
replaces \eq(3.14) of ref.~\cite{relat}.
It should be noted that $m^{ }_\phi = 0$ in the terms shown in \eq\nr{nlo}, 
which will play a role in the following (cf.\ \se\ref{se:X}). 

%%%%%%%%%%%%%%%%%%%%%%%%%%%%%%% FIGURE %%%%%%%%%%%%%%%%%%%%%%%%%%%%%%%%%%%%%%
%
\begin{figure}[t]

\vspace*{-1.0cm}%

%\begin{minipage}[c]{3cm}
\begin{eqnarray*}
&& 
 \hspace*{-0.7cm}
 \GraphLxxiv
 \hspace*{0.75cm}
 \GraphLxxiii
 \hspace*{0.75cm}
 \GraphLxxii
 \hspace*{0.75cm}
 \GraphLxxvi
 \hspace*{0.75cm}
 \GraphLxxv
 \\[1mm] 
&&
 \hspace*{-0.05cm}
 \mbox{(a)}
 \hspace*{2.65cm}
 \mbox{(b)}
 \hspace*{2.45cm}
 \mbox{(c)}
 \hspace*{2.85cm}
 \mbox{(d)}
 \hspace*{2.65cm}
 \mbox{(e)}
\end{eqnarray*}
%\end{minipage}

\vspace*{-0.5cm}%

\caption[a]{\small 
 Examples of processes for right-handed neutrino production:
 (a) a tree-level process producing a massive right-handed neutrino
 (double line) out of a coalescence of a Higgs (dashed line) and a 
 left-handed lepton (solid line); 
 (b) the same process after HTL resummation, generating thermal 
 self-energies or effective vertices (filled blobs);
 (c) processes contributing at the same order as (b), 
 due to exchanges of soft $W^\pm,Z^0,\gamma$ bosons; 
 (d) another channel allowed by the thermal masses generated 
 by HTL resummation (if $m_\phi > M$); 
 (e)~processes contributing at the same order as (d). 
 In the language of \eq\nr{lpm_pre}, the processes (a)--(c) 
 originate from the range $\ko = \omega_1+\omega_2$, 
 $\omega_1 > 0$, $\omega_2 > 0$, whereas (d) and (e) 
 correspond to $\omega_2 = \ko - \omega_1$, $\omega_1 < 0$, 
 where $\omega_1$ is the lepton, $\omega_2$ the Higgs, 
 and $\ko$ the right-handed neutrino energy. 
} 
\la{fig:lpm}
\end{figure}
%
%%%%%%%%%%%%%%%%%%%%%%%%%%%%%%%%%%%%%%%%%%%%%%%%%%%%%%%%%%%%%%%%%%%%%%%%%%%%%

Let us reiterate that even though expressed in a concise form 
in \eq\nr{nlo}, the NLO expression incorporates many types
of physical processes. There are real $2\rightarrow 2$ and $3\rightarrow 1$
reactions that can be assembled into compact expressions, 
given in appendix~A, which could also have been derived from  
a Boltzmann equation. In addition there are ``virtual'' corrections, 
i.e.\ self-energy and vertex insertions into the 
$2\rightarrow 1$ process, given in appendix~B. These can be pictured
as in \fig\ref{fig:lpm}, and include thermal effects represented through the 
HTL effective theory~\cite{htl1,htl2}
(actually the HTL vertex correction vanishes). 
In the NLO computation the self-energy 
corrections appear as insertions rather than in a fully 
resummed form, but they nevertheless suffice to cancel  
soft divergences from the real processes. 

%%%%%%%%%%%%%%%%%%%%%%%%%%%%% SECTION %%%%%%%%%%%%%%%%%%%%%%%%%%%%%%%%%%%%
%
\section{LPM resummation for light-cone kinematics}
\la{se:lpm}

When $M\ll \pi T$, all particles participating in 
$2\to 1$ processes would be essentially ``massless'' from the 
point of view of the thermal motion characterized by the scale $\pi T$, 
were it not that they obtain effective 
thermal masses $\sim gT$ 
through interactions with the other particles in the plasma
(the thermal mass of the right-handed
neutrino is $\sim |h^{ }_\nu| T$
and can be neglected for $|h^{ }_\nu| \ll 1$). 
Typical processes taking place in this situation are depicted
in \fig\ref{fig:lpm}. Instead of $2\rightarrow 1$ coalescence, 
$1\rightarrow 2$ decays of thermal Higgs quasiparticles are the dominant
``tree-level'' process if $gT \gg M$. In addition, however, there
are higher-order scatterings, such as those shown in 
\figs\ref{fig:lpm}(c) and \ref{fig:lpm}(e), which are not suppressed
despite the additional vertices, 
because the exchanged $t$-channel 
gauge boson has soft virtuality $\sim g^2T^2$ 
(it is space-like, and regulated by HTL self-energies).
All these processes need to be summed together, which can 
be achieved through a procedure known as LPM-resummation. 

Given that the ``leading'' particles participating 
in the reaction are ultrarelativistic, 
the kinematics of the process 
amounts to an expansion around the light cone ($\ko\to k$).  
Technically this means that kinematic variables are evaluated
as a power series in mass/energy, so that for instance 
$\ko - k \approx M^2/(2\ko)$. For any fixed $M$, this implies
that only momenta $k \gg M$ are treated consistently. 
However, in practice the breakdown of the framework does not
appear to be dramatic even when this inequality is not strictly
satisfied~\cite{dilepton_lpm}. 

Following ref.~\cite{bb1} but changing the notation slightly, the basic
equations for the LPM resummation  
can be expressed as follows. Let us define
a Hamiltonian 
\be
 \hat{H} \; \equiv \; - \frac{M^2}{2\ko} + 
 \frac{m_\ell^2 - \nabla_\perp^2}{2\omega_1} + 
 \frac{m_\phi^2 - \nabla_\perp^2}{2\omega_2} + i \, V(y) 
 \quad y \equiv |\vec{y}| \equiv |\vec{y}^{ }_\perp| 
 \;, \la{H}
\ee
where $\nabla^{ }_\perp$ is a two-dimensional gradient operating in 
directions 
orthogonal to $\vec{k}$ ($\vec{y}^{ }_\perp\cdot\vec{k} = 0$), 
and $m^{ }_\ell$, $m_\phi$ are the
thermal masses from \eq\nr{mH}.\footnote{%
 The overall sign of $iV$ is a convention and can be 
 reversed by a corresponding sign change in \eq\nr{lpm_pre}. 
 } 
The ``potential'' $V$ plays the 
role of a ``thermal width'', erasing phase coherence from  
a ($\phi$, $\ell$) pair as it propagates through the plasma; its form reads
\be
 V({y}) = \frac{T}{8\pi} \sum_{i=1}^{2} d_i\, g_i^2
 \biggl[ \ln\biggl( \frac{m^{ }_\rmii{D$i$} y}{2} \biggr)
 + \gammaE + K_0 \bigl( m^{ }_\rmii{D$i$} y \bigr) \biggr]
 \;, \la{V}
\ee
where $d_1 \equiv 1$; $d_2 \equiv 3$;  
$K_0$ is a Bessel function; and 
$m^{ }_\rmii{D1}$, $m^{ }_\rmii{D2}$ are electric screening 
masses for the U(1) and SU(2) gauge bosons:
\be
 m_\rmii{D1}^2 = \frac{11}{6} g_1^2 T^2
 \;, \quad
 m_\rmii{D2}^2 = \frac{11}{6} g_2^2 T^2
 \;. \la{mE}
\ee
With the Hamiltonian at hand, we need to solve the inhomogeneous equations
\be
 (\hat{H} + i 0^+)\, g(\vec{y}) \, = \, 
  \delta^{(2)}(\vec{y}) \;, \quad 
 (\hat{H} + i 0^+)\, \vec{f}(\vec{y}) \, = \, 
  -\nabla^{ }_\perp \delta^{(2)}(\vec{y}) 
 \;. \la{Seq}
\ee
Then the LPM-resummed contribution to the correlator 
of \eqs\nr{Pi}, \nr{relation} reads~\cite{bb1}
\ba
 {\im\Pi_\rmii{R}^\rmii{LPM}} % {|h_\nu|^2}
 & \equiv & 
 - \frac{1}{4\pi} 
 \int_{-\infty}^{\infty} \! {\rm d}\omega_1 \, 
 \int_{-\infty}^{\infty} \! {\rm d}\omega_2 \;
 \delta(\ko - \omega_1 - \omega_2) \, 
 \bigl[ 1 - \nF{}(\omega_1) + \nB{}(\omega_2) \bigr] 
 \nn  
 & \times & 
  \, \frac{\ko}{\omega_2} 
 \lim_{\vec{y} \to \vec{0}}
 \biggl\{
   \frac{M^2}{\ko^2} \im\, \bigl[g(\vec{y} )\bigr]  + 
   \frac{1}{\omega_1^2} \im\, \bigl[\nabla_\perp\cdot \vec{f}(\vec{y} )\bigr] 
 \biggr\}
 \;. \la{lpm_pre}
\ea

Apart from the LPM result, hard $2\rightarrow 2$ scatterings
also contribute in the ultrarelativistic regime~\cite{bb2}. In our approach
these originate from the part of the NLO result 
(\se\ref{se:nlo}) which is left over when 
the resummed $1+n \leftrightarrow 2+n$ 
processes, now part of the LPM expression, are subtracted. 
This subtraction is accomplished in \se\ref{se:subtraction}.  

As has been discussed in ref.~\cite{dilepton_lpm}
(following a strategy originally proposed in ref.~\cite{original}), 
the solutions of the 
inhomogeneous equations \nr{Seq} can be reduced to regular solutions
($u^r_\ell$) of 
the corresponding homogeneous equations, with a specific angular quantum
number (denoted by $\ell \in \mathbbm{Z}$). Introducing 
a dimensionless variable $\rho \equiv y\, m^{ }_\rmii{D2}$, 
the homogeneous equation reads
\be
 \biggl[
   -\frac{{\rm d}^2}{{\rm d} \rho^2} + \frac{\ell^2-1/4}{\rho^2}
   + \frac{ M_\rmi{eff}^2 (\omega)  }{m_\rmii{D2}^2} 
   + 2 i\, \frac{\omega (\ko-\omega)}{\ko m_\rmii{D2}^2} 
   \, V\Bigl( \frac{\rho}{m^{ }_\rmii{D2}} \Bigr)
 \biggr] \, u_\ell^r(\rho) \; = \; 0
 \;, \la{url}
\ee
where the particle masses appear through the combination
\be
 M_\rmi{eff}^2 (\omega) \, \equiv \,
 \frac{(\ko-\omega) m_\ell^2}{\ko} + \frac{\omega  m_\phi^2 }{\ko} 
 - \frac{ \omega(\ko - \omega) M^2 }{\ko^2} 
 \;. \la{Meff}
\ee
Choosing normalization such that the small-$\rho$ asymptotics reads
\be
 u_\ell^{r}(\rho)
 \, = \, \rho^{1/2 + |\ell|} \, \bigl[ 1 + \rmO(\rho^2) \bigr]
 \;, \la{url_asymptotics}
\ee
\eq\nr{lpm_pre} can then be re-written as
\ba
 {\im\Pi_\rmii{R}^\rmii{LPM}} % {|h_\nu|^2}
 & = & 
 - \frac{1}{\pi^2} 
 \int_{-\infty}^{\infty} \! {\rm d}\omega \;
 \bigl[ 1 - \nF{}(\omega) + \nB{}(\ko - \omega) \bigr] 
 \nn  
 & \times & 
 \int_0^\infty \! {\rm d}\rho \, 
 \biggl[
   \frac{\omega M^2}{4\ko^2}
   \im \biggl\{ \frac{ 1 }{[u_0^r(\rho)]^2} \biggr\} 
 +
   \frac{m_\rmii{D2}^2}{\omega}
   \im \biggl\{ \frac{1}{[u_1^r(\rho)]^2} \biggr\} 
 \biggr]
 \;. \la{lpm}
\ea
We have checked numerically that this agrees with the results
of ref.~\cite{bb1}.

%%%%%%%%%%%%%%%%%%%%%%%%%%%%% SECTION %%%%%%%%%%%%%%%%%%%%%%%%%%%%%%%%%
%
\section{Subtraction of a $2\leftrightarrow 1$ part from 
 the NLO expression}
\la{se:subtraction}

The NLO expression of \se\ref{se:nlo} contains 
$2 \rightarrow 1$, $2\rightarrow 2$, and $3\rightarrow 1$ 
processes (cf.\ appendix~A). 
The LPM resummation of \se\ref{se:lpm} treats $2+n \leftrightarrow 1+n$
processes to all orders. However, it does nothing to 
$2\rightarrow 2$ and $3\rightarrow 1$ processes. Therefore, 
the $2\rightarrow 2$ and $3\rightarrow 1$ processes of the NLO expression
need to be added to the LPM result~\cite{bb2}. But in order 
to avoid double counting in doing so, the $2 \rightarrow 1$ part
needs first to be subtracted from the NLO expression.

Of course, the $2 \rightarrow 1$ part cannot be subtracted 
from the NLO expression as such, because virtual corrections make it infrared 
divergent. Yet it is only a particular subpart of the virtual corrections
to the $2 \rightarrow 1$ processes, given by HTL effects, which
play a role in the LPM resummation. It turns out that if 
we carry out HTL resummation also in the real $2\rightarrow 2$ processes
dominated by soft momentum transfer, then the $2\rightarrow 1$
and $2\rightarrow 2$ processes are {\em separately} infrared finite. Then we 
can subtract the HTL-induced $2\rightarrow 1$ part from the NLO expression, 
and add the remainder to the LPM result.

It is important to note that since in the NLO part of \eq\nr{nlo} all internal 
particles are massless, we need to set $m_\phi \to 0$ in the corresponding 
terms 
of the present section. This does not lead to any divergences, however there
is a certain endpoint sensitivity related to Bose-enhanced 
Higgs bosons which needs to be treated with care
(cf.\ \se\ref{se:X}).\footnote{%
 A different logic for determining the subtraction term
 was presented for 
 the case of dilepton production from hot QCD
 in ref.~\cite{dilepton_lpm}. 
 Applying the same logic here would reproduce
 \eq\nr{virtual_2}, apart from the term $T/\ko$, whose  
 sign depends on the ordering of taking $m_\phi/M\to 0$
 and $M/k \to 0$ in terms of $\rmO(m_\ell^2)$.  
 }

We proceed in steps. First, the $2\to 2$ corrections need to be modified
such that soft momentum transfer is regulated by HTL resummation. 
In the NLO expression of \se\ref{se:nlo}, all the correct HTL structures 
do appear, albeit as ``insertions'' rather than in a resummed form. 
As is shown in appendix~C (cf.\ \eq\nr{delta_4}), 
this means that the 
soft contribution to $2\rightarrow 2$ processes appears as
\be
 \im \Pi_\rmii{R}^\rmii{HTL,ins,cut} \; \equiv \;
 \frac{m_\ell^2}{4\pi}
 \, \Bigl[
   \nB{}(\ko) + \fr12 
 \Bigr] \, 
 \Bigl(
   \ln\frac{2\Lambda}{\lambda} - 1 
 \Bigr)
 \;, \la{htl_cut_ins}
\ee 
where $\Lambda$ is an ultraviolet cutoff ($g T \ll \Lambda \ll \pi T$)
and $\lambda$ is an infrared regulator. If the same processes are treated
with full HTL resummation, the result gets modified into 
(cf.\ \eq\nr{delta_2})
\be
 \im \Pi_\rmii{R}^\rmii{HTL,full,cut} \; \equiv \;
 \frac{m_\ell^2}{4\pi}
 \, \Bigl[
   \nB{}(\ko) + \fr12 
 \Bigr] \, 
 \Bigl(
   \ln\frac{2\Lambda}{m^{ }_\ell} - 1 
 \Bigr)
 \;. \la{htl_cut_full}
\ee 
These results will be needed presently. 

Consider then the $2\rightarrow 1$ part of the NLO 
expression (cf.\ appendix~B). Carrying out the velocity 
integral as well as the integral over the angles between 
$\vec{p}$ and $\vec{k} $ in \eq\nr{HTLpole}, and 
taking the infrared regulator $\lambda\to 0$ wherever possible, 
\eq\nr{HTLpole} can be expressed as
\ba
 & & \hspace*{-1.5cm} 
 \im\Pi_\rmii{R}^\rmii{HTL,ins,pole} \; \equiv \;
 \frac{1}{8\pi k}
 \int_{\km - \frac{m_\ell^2}{4\km}}^{\kp - \frac{m_\ell^2}{4\kp}}
 \, \frac{{\rm d} p \, p}{ \epsilon^{ }_\ell} \, 
 (M^2 + m_\ell^2) \, 
 \Bigl[ 1 - \nF{}(\epsilon^{ }_\ell) + \nB{}(\ko - \epsilon^{ }_\ell)  \Bigr]
 \nn & + & 
 \frac{m_\ell^2}{8\pi k}
 \int_{\km}^{\kp}
 \!\! {\rm d} p \, 
 \biggl[ - \frac{\ko}{p}  + 
 \frac{M^2}{2 p^2}\biggl( 1 - \ln\frac{2p}{\lambda}\biggr) \biggr] \, 
 \Bigl[ 1 - \nF{}(p) + \nB{}(\ko - p) \Bigr]
 \;. \la{HTLpole2}
\ea
Here $\epsilon^{ }_\ell \equiv \sqrt{p^2 + m_\ell^2}$
and it is understood that we expand to $\rmO(m_\ell^2)$
after the computation. 
The integration ranges in \eq\nr{HTLpole2} originate from 
$\delta(\ko - \epsilon^{ }_\ell - |\vec{p}-\vec{k}|)$ in the terms 
where the next-to-leading order in $m_\ell^2$ is needed, and from
$\delta(\ko - p - |\vec{p}-\vec{k}|)$ otherwise.

The first line of \eq\nr{HTLpole2} is readily integrated by taking
$\epsilon_\ell$ as an integration variable. Expanding subsequently 
in $m_\ell^2$ and taking also the limit $M \ll k$ leads to 
\be
  \delta_1 \im\Pi_\rmii{R}^\rmii{HTL,ins,pole}
  \; \equiv \; 
  \frac{m_\ell^2}{8\pi}
  \biggl\{
    \frac{T}{\ko} + 
   \int_0^{\ko} \! \frac{{\rm d}\omega}{\ko} \, 
   \Bigl[ 
     \nB{}(\ko - \omega) - \nB{}(\ko) - \nF{}(\omega) + \nF{}(0) 
   \Bigr] 
  \biggr\}
  \;, \la{htl_pole_1}
\ee
where the term of $\rmO(m_\ell^0)$ was omitted (it will be added 
separately later on). In order to integrate the term $-\ko/p$ on 
the second line of \eq\nr{HTLpole2}, we add and subtract the value of 
$
 1 - \nF{}(p) + \nB{}(\ko - p)
$
at $p=0$ and take subsequently $M \ll k$, 
producing 
\ba
  \delta_2 \im\Pi_\rmii{R}^\rmii{HTL,ins,pole}
  & \equiv &  
  \frac{m_\ell^2}{8\pi}
  \biggl\{
   \Bigl[\nB{}(\ko) + \fr12 \Bigr] \ln\Bigl( \frac{M^2}{4\ko^2} \Bigr)
 \nn & - & 
   \int_0^{\ko} \! \frac{{\rm d}\omega}{\omega} \, 
   \Bigl[ 
     \nB{}(\ko - \omega) - \nB{}(\ko) - \nF{}(\omega) + \nF{}(0) 
   \Bigr] 
  \biggr\}
  \;. \la{htl_pole_2}
\ea
For the last part of \eq\nr{HTLpole2}, 
we carry out a similar addition-subtraction step, 
and note that with the subtracted weight, the integral is only 
logarithmically sensitive to the lower bound. This yields a contribution 
of $\rmO((m_\ell^2 M^2 / \ko^2) \ln^2 (M^2/2\lambda\ko)  ) 
\sim \rmO(g^4 T^2)$ which is omitted from the HTL consideration
(the logarithmic divergence cancels against real corrections). 
The other part is easily integrated, 
and a substitution at the lower bound
gives % a contribution of $\rmO(g^2 T^2)$:
\be
  \delta_3 \im\Pi_\rmii{R}^\rmii{HTL,ins,pole}
  \; \equiv \;   
  \frac{m_\ell^2}{8\pi}
   \Bigl[\nB{}(\ko) + \fr12 \Bigr]
   \ln\Bigl( \frac{4\lambda^2\ko^2}{M^4} \Bigr)
  \;. \la{htl_pole_3}
\ee 

We can now define the HTL-induced 
$2\to 1$ part of the NLO result. We write this as 
\be
 \im \Pi_\rmii{R}^\rmii{$\Delta$\VIRTUAL} \; \equiv \; 
  \im \Pi_\rmii{R}^\rmii{$\Delta$\VIRTUAL,$m_\ell^0$}
 + 
   \im \Pi_\rmii{R}^\rmii{$\Delta$\VIRTUAL,$m_\ell^2$} 
 \;. \la{virtual_sum}
\ee
The term of $\rmO(m_\ell^0)$ was omitted 
above; it needs to be computed in the presence of $m_\phi > 0$ 
like \eq\nr{tree}. In fact the result is a limit of \eq\nr{tree} but 
for the kinematics 
$M^2  \ll \mathop{\mbox{min}}( \ko^2, \ko m^{ }_\phi, \ko T )$
assumed in the LPM computation: 
\ba
 { \im\Pi_\rmii{R}^\rmii{$\Delta$\VIRTUAL,$m_\ell^0$} } % {|h_\nu|^2}
 & \equiv & 
 \frac{(M^2 - m_\phi^2)T}{8\pi\ko}
 \ln\left\{
   \frac{\sinh\Bigl( \frac{\ko}{2T} \Bigr)
   \cosh\Bigl[ \frac{\ko}{2T} \Bigl(1 -  \frac{m_\phi^2}{M^2} \Bigr)\Bigr] }
  {\sinh \Bigl( \frac{\ko m_\phi^2}{2T M^2} \Bigr) }
 \right\}
 \;. \la{virtual_0}
\ea
The correction of $\rmO(m_\ell^2)$, in turn, 
is a sum of \eqs\nr{htl_pole_1}--\nr{htl_pole_3}
and the cut contribution from \eq\nr{htl_cut_ins} 
(reshuffled here from the original $2\to 2$ corrections), from 
which the HTL-resummed
cut of \eq\nr{htl_cut_full} is subtracted 
(since this should now appear as a part of $2\to 2$ corrections):
\ba
 { \im\Pi_\rmii{R}^\rmii{$\Delta$\VIRTUAL,$m_\ell^2$} } 
 & \equiv & 
  ( \delta_1 
  + \delta_2 
  + \delta_3 ) \im\Pi_\rmii{R}^\rmii{HTL,ins,pole}
  +  \im \Pi_\rmii{R}^\rmii{HTL,ins,cut} 
  -  \im \Pi_\rmii{R}^\rmii{HTL,full,cut}
 \nn[3mm] 
 & = & 
   \frac{m_\ell^2}{8\pi}
  \biggl\{
  \Bigl[\nB{}(\ko) + \fr12 \Bigr]
  \ln\Bigl( \frac{m_\ell^2}{M^2} \Bigr) + 
  \frac{T}{\ko} 
 \nn & + & 
   \int_0^{\ko} \! {\rm d}\omega \, 
  \biggl( \frac{1}{\ko} - \frac{1}{\omega} \biggr) \, 
   \Bigl[ 
     \nB{}(\ko - \omega) - \nB{}(\ko) - \nF{}(\omega) + \nF{}(0) 
   \Bigr] 
  \biggr\}
 \;. \hspace*{5mm} \la{virtual_2}
\ea
Note that both the ultraviolet cutoff $\Lambda$ and the infrared
regulator $\lambda$ have dropped out here. 
The expression in \eq\nr{virtual_2}
needs to be subtracted from the NLO computation as will be discussed
in more detail in \se\ref{se:interpolate}, 
but before doing so we need to clarify 
one feature of virtual $2\to 1$ corrections that goes beyond the HTL limit. 

%%%%%%%%%%%%%%%%%%%%%%%%%%%%% SECTION %%%%%%%%%%%%%%%%%%%%%%%%%%%%%%%%%%%%
%
\section{Issues related to contributions from soft Higgs bosons}
\la{se:X}
  
It was mentioned in \se\ref{se:nlo} that, even though the Higgs mass needs
to be thermally resummed when $M \lsim g^{1/2}T$ as done in the LO result 
of \eq\nr{tree}, the Higgs mass was kept at zero in the NLO contribution of 
\eq\nr{nlo}. This does not cause the NLO contribution to diverge; 
nevertheless, as we now argue, its value is not physically correct 
for $M \ll g^{1/2} T$. 

The reason for the problem is that there are
parts in the NLO contribution which include two Bose factors. 
This may lead
to an enhancement $\sim T^2/\epsilon_\phi^2$, where $\epsilon^{ }_\phi$
is the Higgs energy. If the Higgs were massless, its energy could be 
as small as $\sim M^2/(4 k)$, whereby we may obtain  
a contribution $\sim 4k T^2/ M^2$ from 
$\int \! {\rm d}\epsilon^{ }_\phi$. Therefore, even 
if there were a prefactor $M^2$, a finite result could be 
left over. Such an expression is not correct, however, because in the presence
of a mass, $\epsilon^{ }_\phi \ge m^{ }_\phi$, and there can be  
no divergence at $M/k \to 0$. 

There are only two structures in \eq\nr{nlo} 
which suffer from this problem, namely the master spectral 
functions $\rho^\rmii{$T$}_{{\mathcal{I}}^{ }_\rmii{h}}$ 
and $\rho^\rmii{$T$}_{{\mathcal{I}}^{ }_\rmii{j}}$ which, 
according to refs.~\cite{relat,master},  
have a finite limit for $M/T \to 0$ even though they 
contain no HTL structures.
Employing the same notation as in appendices~A and B, their 
contributions to \eq\nr{nlo} can formally be expressed as 
\ba
 && \hspace*{-1.5cm}
  \nF{}(\ko) 
  \Bigl( 
      \rho^\rmii{$T$}_{{\mathcal{I}}^{ }_\rmii{j}}
    - 2 
       \rho^\rmii{$T$}_{{\mathcal{I}}^{ }_\rmii{h}}
   \Bigr) 
  \nn 
  & = & 
 \int \! {\rm d}\Omega^{ }_{2\to 1} \;
 \nF{}(p_1) \, \nB{}(p_2) \, 
    \Tint{ Q } \frac{M^2}{Q^2(Q-P^{ }_2)^2}
    \, 
  \biggl[
  1 
    + 
  \frac{M^2}{(Q-K)^2} \biggr] 
    \, \biggr|^{ }_{p^{ }_{2n} = - i p_2,\, k_n = - i \ko}
 \nn 
 & + & 
 \int \! {\rm d}\Omega^{ }_{2\to 2} \; 
 \biggl\{\; 
  \nB{}(p_1) \, \nB{}(p_2) \, \bigl[1 - \nF{}(k_2)\bigr]
 \, \biggl[ - \frac{ M^2 (M^2-t)}{st} \biggr]
 \nn & & \hspace*{1.6cm} + \, 
   \nF{}(p_1) \, \nB{}(p_2) \, \bigl[1 + \nB{}(k_2)\bigr]
 \, \biggl[
  \frac{ M^2(M^2 - s)}{st}
+ \frac{ M^2(M^2 - u)}{ut}
 \biggr]
 \; \biggr\} 
 \nn 
 & + & 
 \int \! {\rm d}\Omega^{ }_{3\to 1} \; 
 \biggl\{\; 
  \nF{}(p_1) \, \nB{}(p_2) \, \nB{}(p_3) 
 \, \biggl[ 
 \frac{ M^2 (M^2 - s_{12})}
 {s_{12} s_{23} }
 \biggr]
 \; \biggr\} 
 \;. \la{problem}
\ea

Let us illustrate the issue with the simplest term, the self-energy 
correction ($P^{ }_2 \equiv (p^{ }_{2n},\vec{p}^{ }_2)$)
\be
 \phi(p_2) \;\equiv\; \Tint{ Q } \frac{1}{Q^2(Q-P^{ }_2)^2}
  \, \biggr|^{ }_{p^{ }_{2n} = - i p_2}
 \;. 
\ee
In naive power counting this would be of $\rmO(1)$
and contains {\em no}
HTL structures $\sim T^2$~\cite{brpi}. However, taken literally, 
$\phi(p_2)$ is infrared divergent. If we insert a regulator, 
$1/(Q-P^{ }_2)^2 \to 1/[(Q-P^{ }_2)^2+\lambda^2]$, and note 
that the Bose-enhanced singularities at $q=0$ and $q=p_2$ cancel against
real corrections (cf.\ ref.~\cite{relat}, \eq(B.70) ff), the Bose-enhanced
part of the result can be written as 
(away from singular points and after a partial 
cancellation against real corrections as well as a substitution
$q \to q+p_2$ in one of the terms)
\be
 \phi(p_2) \simeq \frac{T}{(4\pi)^2 p_2}
 \int_0^\infty  \! \frac{{\rm d}q}{q} \ln \biggl| \frac{p+q}{p-q} \biggr| 
 = \frac{T}{32 p_2}
 \;. \la{Top}
\ee
Since the answer grows only linearly in $T$, 
it is not part of HTLs. Inserting the result into the $2\to 1$
phase space integral yields subsequently 
\be
 \nF{}(\ko) \, \delta \rho^\rmii{$T$}_{{\mathcal{I}}^{ }_\rmii{h}}  
 \simeq
 - \frac{\pi M^2 T}{32 (4\pi)^2 k} \int_{\km}^{\kp} \! 
 \frac{ {\rm d}p_2 }{p_2} \, 
 \nB{}(p_2) \nF{}(\ko - p_2)
 \simeq 
 - \frac{\pi M^2 T^2 \nF{}(\ko) }{32(4\pi)^2 k\, \km}
 \;, \la{appro_wrong}
\ee
where we only kept the Bose-enhanced 
contribution from the lower edge of the integration range.
Given that $\km \approx M^2 / (4k)$ for $M \ll k$, the factor $M^2$
is seen to cancel out, leaving over a finite contribution. 

This finite contribution is not ``correct'', however. Indeed, had we had
the Higgs mass in the $2\to 1$ phase space integral, the minimal Higgs energy
would have been 
 $\epsilon_\phi^\rmi{min} = \km + {m_\phi^2}/({4\km})$
for $M > m^{ }_\phi$, 
and 
 $\epsilon_\phi^\rmi{min} = \kp + {m_\phi^2}/({4\kp})$
for $M < m^{ }_\phi$.
Obviously $\epsilon_\phi^\rmi{min} \ge m^{ }_\phi$ for any $M$, and 
to a good approximation $\epsilon_\phi^\rmi{min} \approx \ko$ for 
$M \,\lsim\, m^{ }_\phi$. In $2\to 2$ scatterings
smaller values can appear, but in any case 
$\epsilon_\phi^\rmi{min} \ge m^{ }_\phi$ is always satisfied. 

In order to account for these features properly, the NLO computations 
of refs.~\cite{relat,master}
should be repeated with $m^{ }_\phi > 0$. 
This is a hard task and goes beyond the scope of the present study. 
Here, we rather resort to a phenomenological interpolation which has
the correct limiting values for $M \gg g^{1/2} T$ and $M \ll g^{1/2}T$.

According to $2\to 1$ kinematics
with $\epsilon_\phi^\rmi{min} = \km + {m_\phi^2}/({4\km})$
for $M > m^{ }_\phi$, 
the phase space employed in the NLO
result of \eq\nr{nlo}, as illustrated 
in \eq\nr{appro_wrong}, is physically correct
only if $\km \gg m_\phi^2/(4\km)$. For $M \ll k$ this corresponds to 
$m^{ }_\phi \ll M^2 / (2 k)$. We need to ``switch off'' 
the incorrect contributions as soon as this inequality is not satisfied. 
This can be achieved by defining 
\be
 \im \Pi_\rmii{R}^\rmii{NLO,$\phi$} \; \equiv \; 
 \im \Pi_\rmii{R}^\rmii{NLO} - 
 \Theta\Bigl( \frac{ 2 k m_\phi }{M^2} - 1 \Bigr) 
 (g_1^2 + 3 g_2^2) 
 \bigl( 
    \rho^\rmii{$T$}_{{\mathcal{I}}^{ }_\rmii{j}}
    - 2 
    \rho^\rmii{$T$}_{{\mathcal{I}}^{ }_\rmii{h}}
 \bigr)
 \;, \la{nlo_phi}
\ee
where $\Theta$ is a smoothed step function, for instance
$\Theta(x) \equiv [1 + \tanh(2x)]/2$. Obviously, the recipe 
is purely phenomenological in the intermediate range 
$ M \sim \sqrt{2 k m_\phi} $, but it does have the 
correct limiting values on both sides of this range. 
For illustration of the numerical (un)importance of the precise
choice made, 
we also consider an implementation with another width
of the switch-off region, namely
\be
 \im \tilde{\Pi}_\rmii{R}^\rmii{NLO,$\phi$} \; \equiv \; 
 \im \Pi_\rmii{R}^\rmii{NLO} - 
 \Theta\biggl( \frac{ { m_\phi^2 }/ {4 \km} - \km}{T} \biggr) 
 (g_1^2 + 3 g_2^2) 
 \bigl( 
    \rho^\rmii{$T$}_{{\mathcal{I}}^{ }_\rmii{j}}
    - 2 
    \rho^\rmii{$T$}_{{\mathcal{I}}^{ }_\rmii{h}}
 \bigr)
 \;. \la{nlo_phi_tilde}
\ee
The difference with respect to \eq\nr{nlo_phi} is plotted as a grey band
in \figs\ref{fig:imRESUM}(right) and \ref{fig:spectraRESUM}(left), 
and is practically invisible in 
the final result shown in \fig\ref{fig:spectraRESUM}(left). 

%%%%%%%%%%%%%%%%%%%%%%%%%%%%% SECTION %%%%%%%%%%%%%%%%%%%%%%%%%%%%%%%%%%%%
%
\section{Putting together the NLO and LPM results}
\la{se:interpolate}

After the considerations of \ses\ref{se:subtraction} and \ref{se:X}, 
we are in a position
to subtract the HTL-induced $2\to 1$ part from the NLO expression. The 
remainder, 
incorporating 
HTL-resummed $2\to 2$ processes, 
subleading corrections to $2\to 1$ processes,
and $3\to 1$ processes (which do not contribute when $M/T\to 0$), 
is defined as 
\be
 \im \Pi_\rmii{R}^\rmii{$\Delta$\REAL} \; \equiv \; 
 \im \Pi_\rmii{R}^\rmii{NLO,$\phi$} - \im \Pi_\rmii{R}^\rmii{$\Delta$\VIRTUAL}
 \;, \la{delta_real}
\ee
where $ \im \Pi_\rmii{R}^\rmii{NLO,$\phi$} $ is from 
\eq\nr{nlo_phi} and 
$\im \Pi_\rmii{R}^\rmii{$\Delta$\VIRTUAL}$ is from 
\eqs\nr{virtual_sum}--\nr{virtual_2}.

Returning briefly to $\im \Pi_\rmii{R}^\rmii{$\Delta$\VIRTUAL}$, 
we note that 
the first term on the right-hand side of \eq\nr{virtual_2} 
is logarithmically divergent for $M^2 \to 0$. The 
divergence also appears in the NLO result, and is removed by 
the subtraction in \eq\nr{delta_real}.
This cancellation constitutes a crosscheck of our 
computation. 
 The logarithmic dependence on $m^{ }_\ell$ in \eq\nr{virtual_2}
 can also be compared
 with literature: our result agrees with ref.~\cite{bb2}, after
 noting that $\nB{}(\ko)+ \fr12$ can be expressed as 
 $\nB{}(\ko)/[2 \nF{}(\ko)]$ and that the discussion in ref.~\cite{bb2}
 concerns the integrand of \eq\nr{gamma}.

Having subtracted the $\Delta$\VIRTUAL\ part, 
\eq\nr{delta_real} can be added
to the LPM expression of \se\ref{se:lpm}, 
which leads to our final result: 
\be
 \im\Pi_\rmii{R}^{ } \; \equiv \; 
 \im\Pi_\rmii{R}^\rmii{LPM} + 
 \im\Pi_\rmii{R}^\rmii{$\Delta$\REAL} 
 \;. \la{final}
\ee
How accurate this expression is, depends on 
the regime considered: in the relativistic regime $M\gsim \pi T$
relative errors are of $\rmO(g^3)$, whereas in the ultrarelativistic 
regime $M \lsim gT$ relative errors are likely to be suppressed
by $\rmO(g)$~\cite{km,gm}. In between there is a regime in which the 
expression is not consistent even at leading order, 
as has been discussed in \se\ref{se:X}.

%%%%%%%%%%%%%%%%%%%%%%%%%%%%%%%%% FIGURE %%%%%%%%%%%%%%%%%%%%%%%%%%%%%%%%%
\begin{figure}[t]

%\vspace*{-3cm}

\centerline{%
 \epsfysize=7.5cm\epsfbox{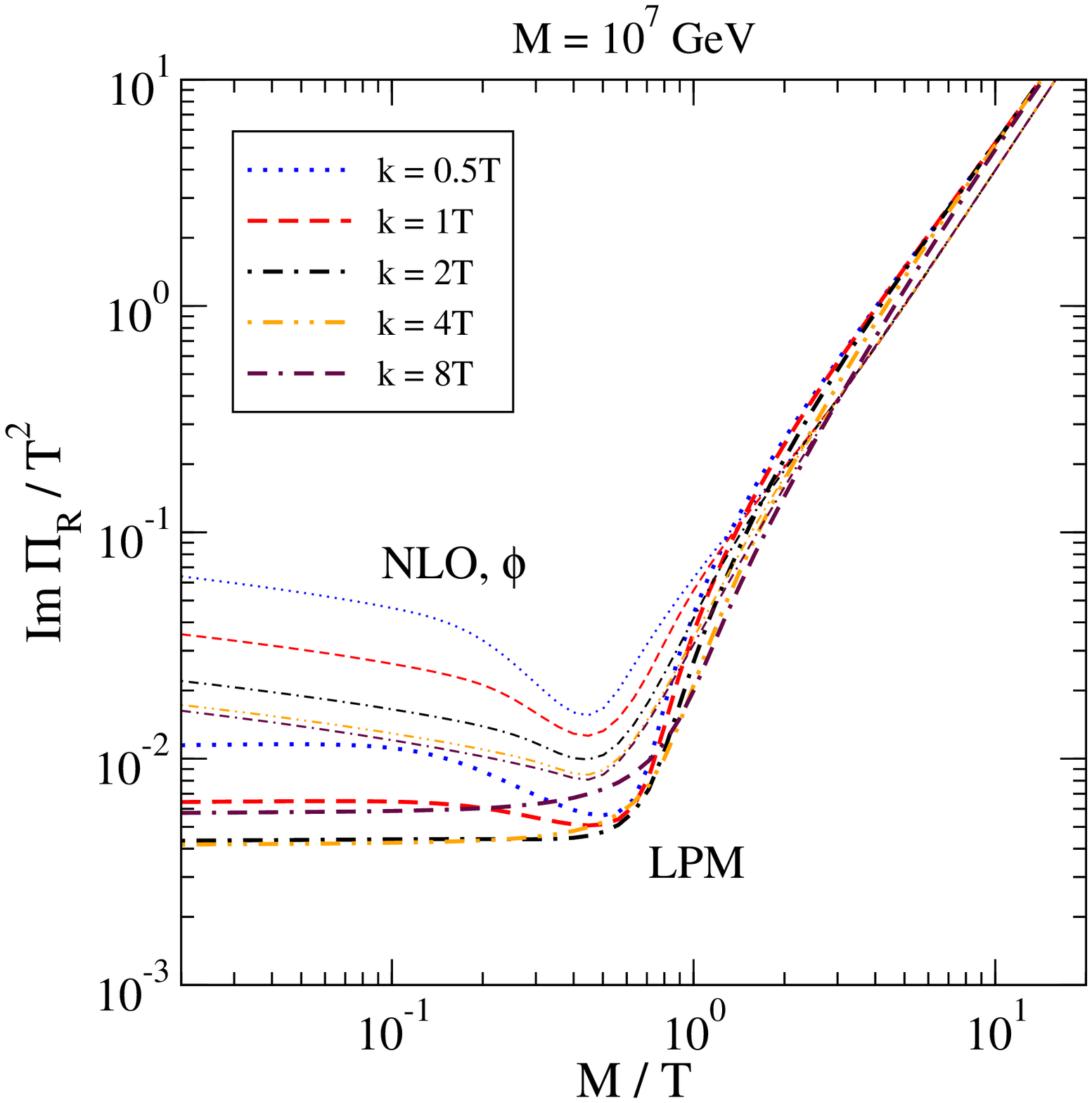}%
~~~\epsfysize=7.5cm\epsfbox{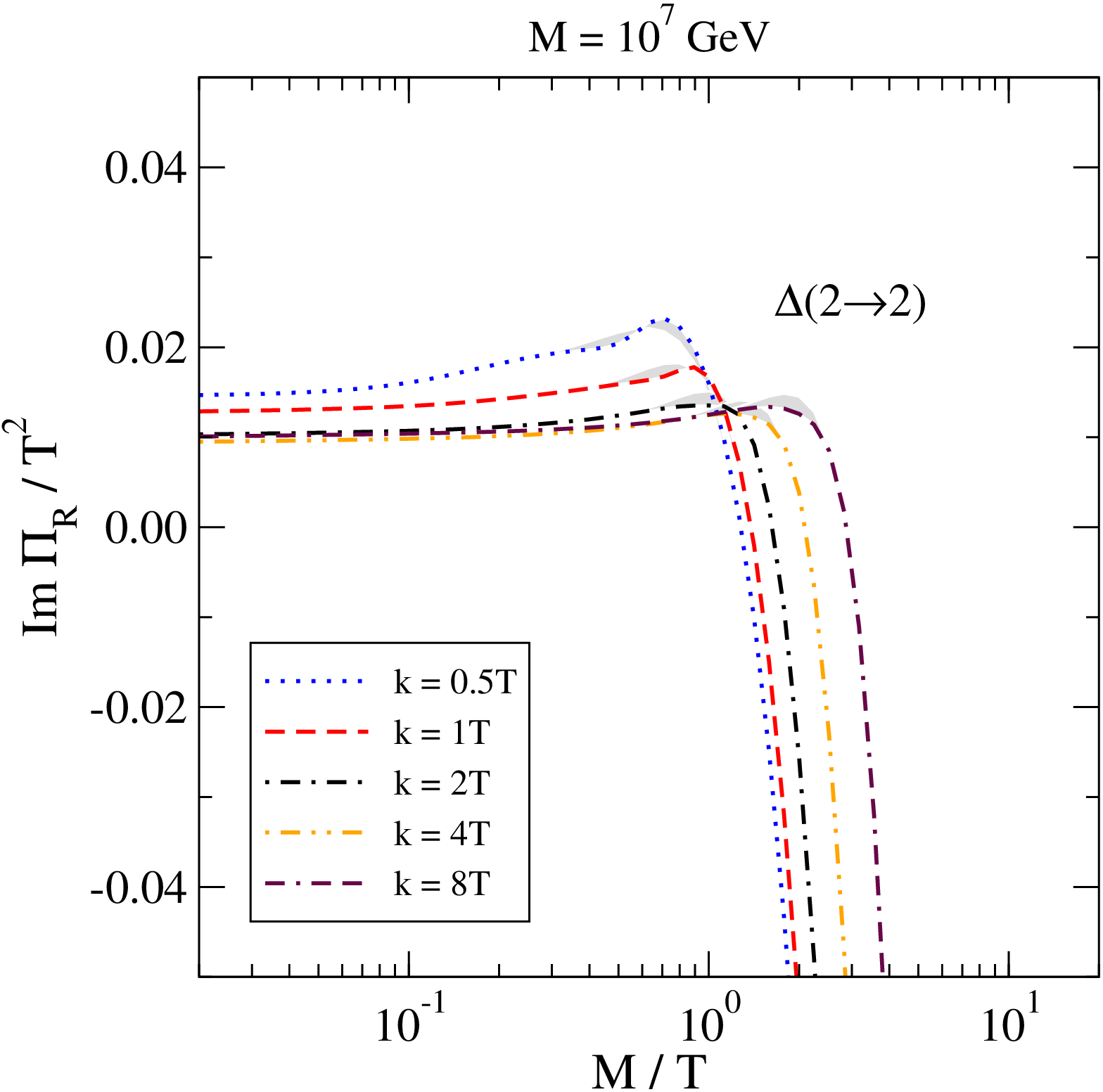}
}

\caption[a]{\small
Left: NLO correlator from \eq\nr{nlo_phi} (thin lines) and
the LPM one from \eq\nr{lpm}
(thick lines). Right: Correction of 
the LPM result through $\Delta$\REAL\  
from \eq\nr{delta_real}. 
The couplings and the renormalization
scale are fixed as specified in appendix~D. 
}

\la{fig:imRESUM}
\end{figure}
%%%%%%%%%%%%%%%%%%%%%%%%%%%%%%%%%%%%%%%%%%%%%%%%%%%%%%%%%%%%%%%%%%%%%%%%%%%

%%%%%%%%%%%%%%%%%%%%%%%%%%%%%%%%% FIGURE %%%%%%%%%%%%%%%%%%%%%%%%%%%%%%%%%
\begin{figure}[t]

%\vspace*{-3cm}

\centerline{%
\epsfysize=7.5cm\epsfbox{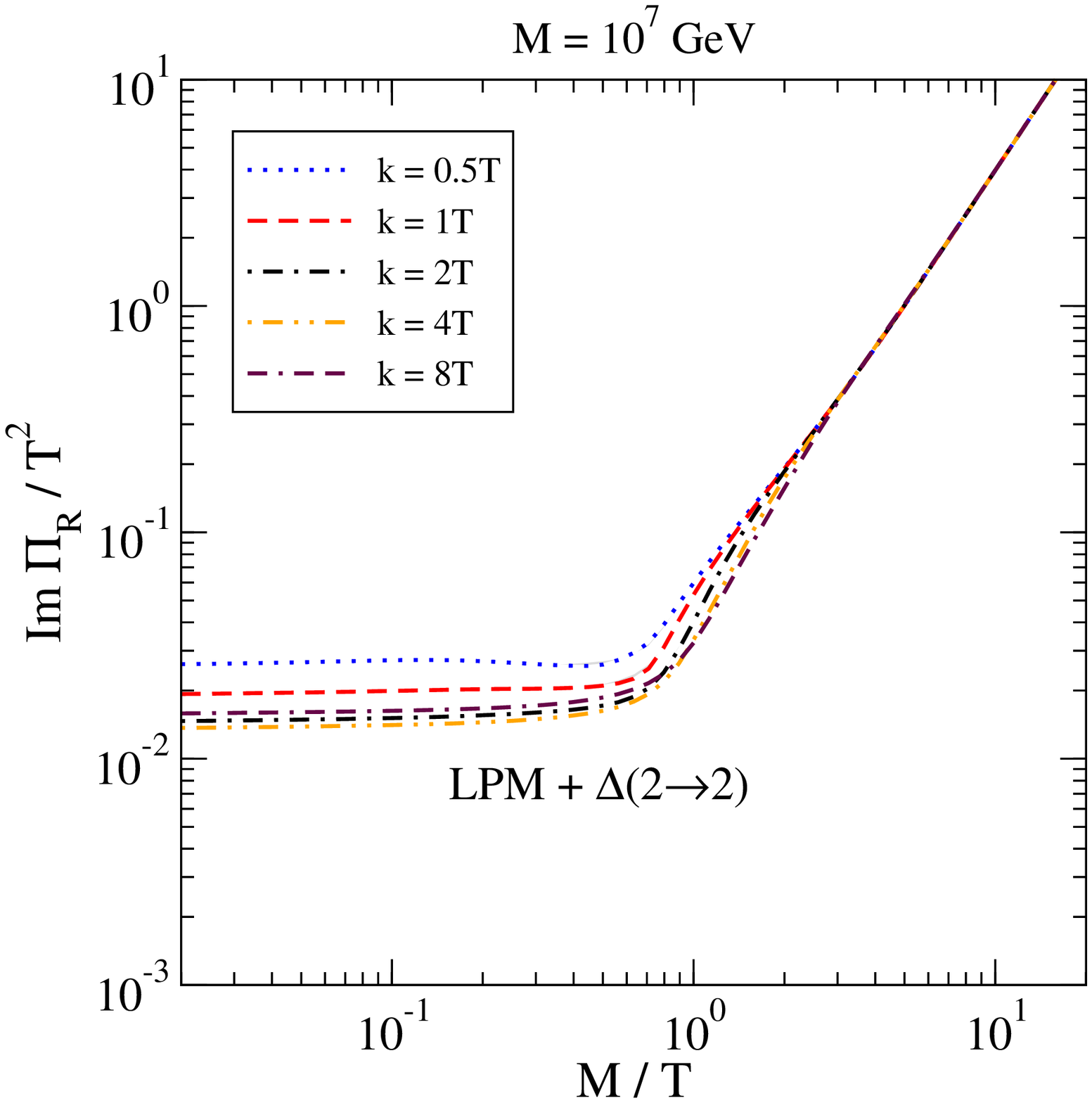}%
~~~\epsfysize=7.5cm\epsfbox{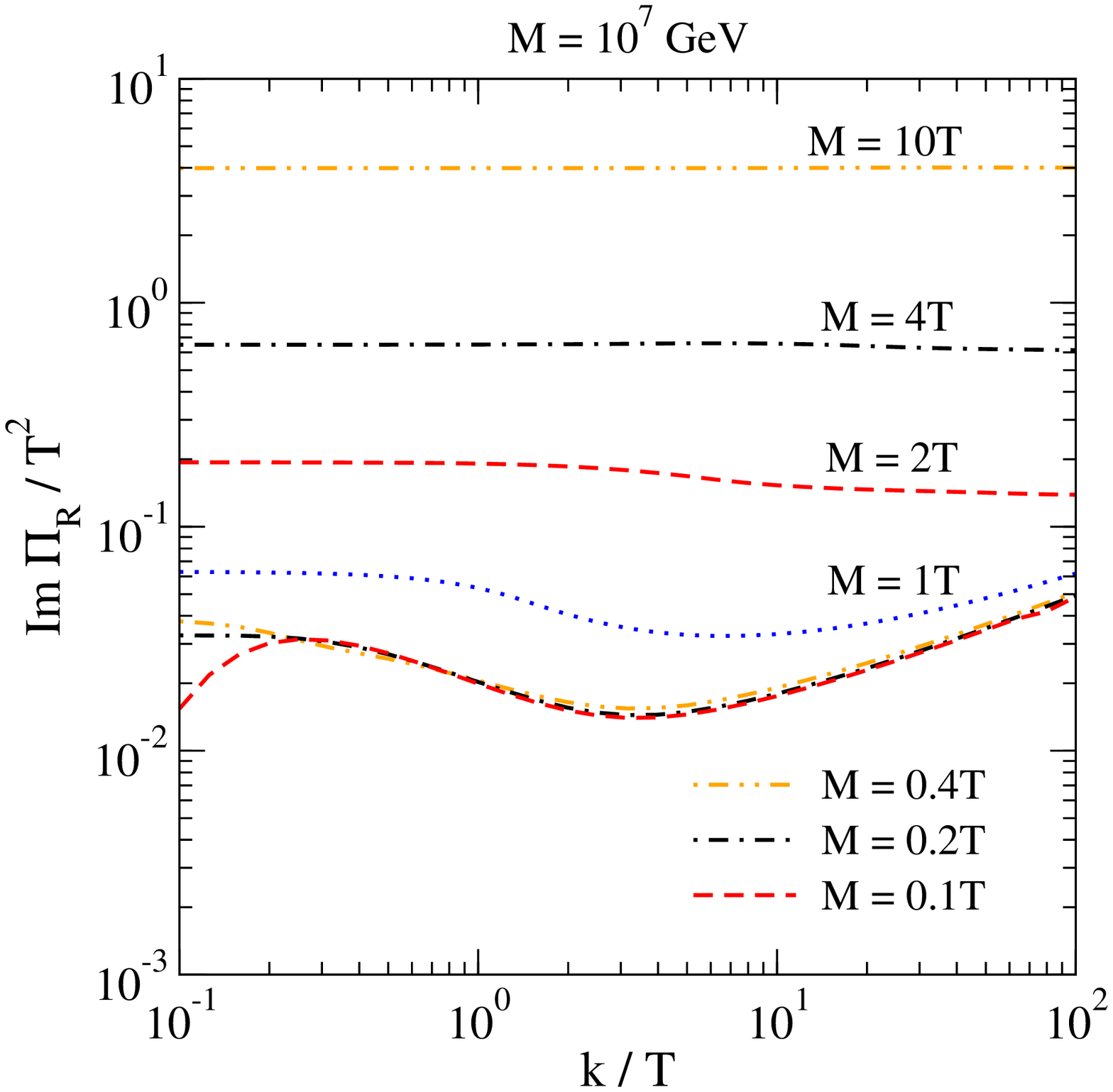}
}

\caption[a]{\small
Left: Final results for the correlator in \eq\nr{final}. 
Right: The same results as a function of $k/T$.
The couplings and the renormalization scale are fixed as
specified in appendix~D.
}

\la{fig:spectraRESUM}
\end{figure}
%%%%%%%%%%%%%%%%%%%%%%%%%%%%%%%%%%%%%%%%%%%%%%%%%%%%%%%%%%%%%%%%%%%%%%%%%%%

%%%%%%%%%%%%%%%%%%%%%%%%%%%%%%%%% FIGURE %%%%%%%%%%%%%%%%%%%%%%%%%%%%%%%%%
\begin{figure}[t]

%\vspace*{-3cm}

\centerline{%
\epsfysize=7.5cm\epsfbox{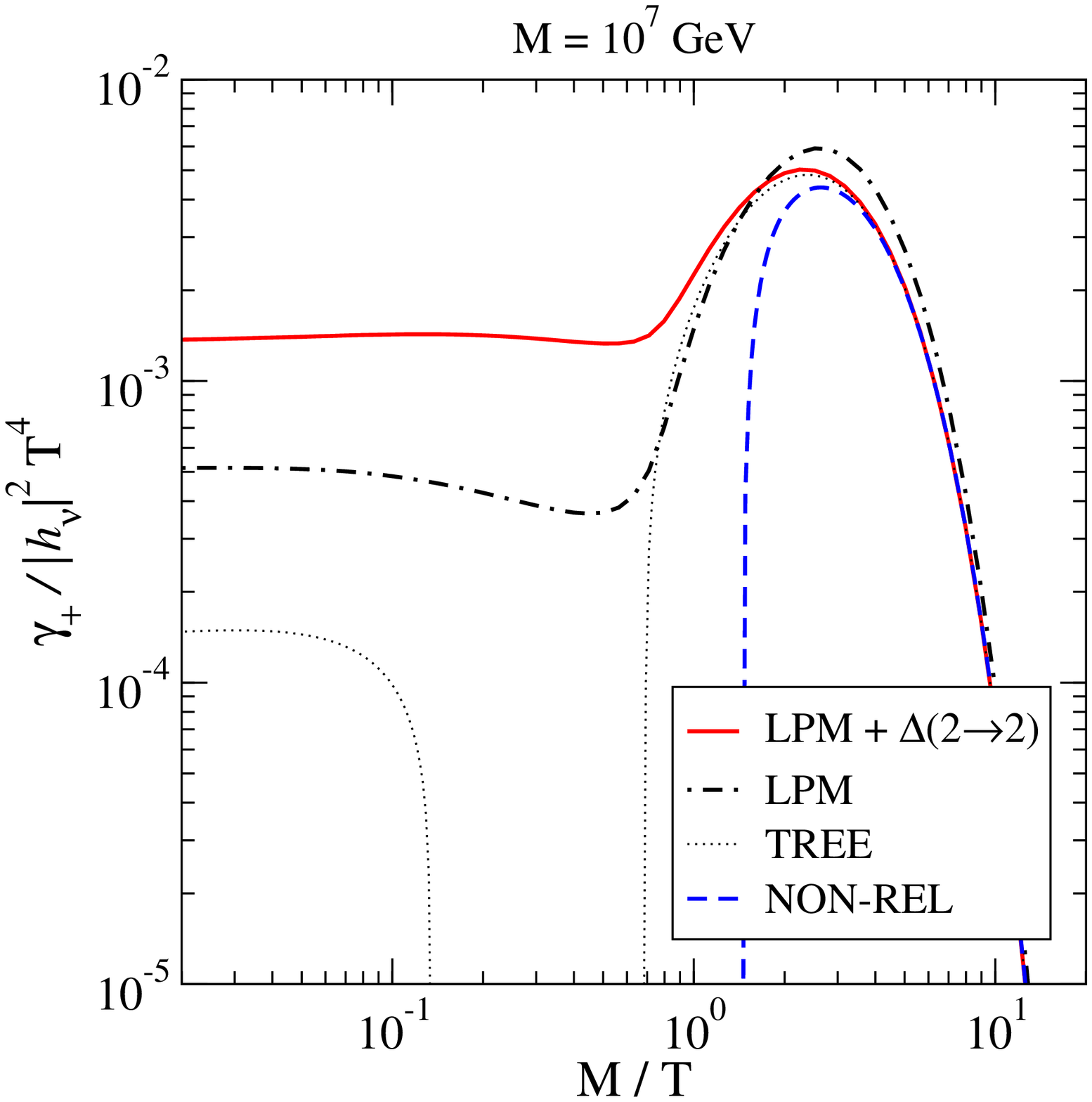}%
 ~~~\epsfysize=7.5cm\epsfbox{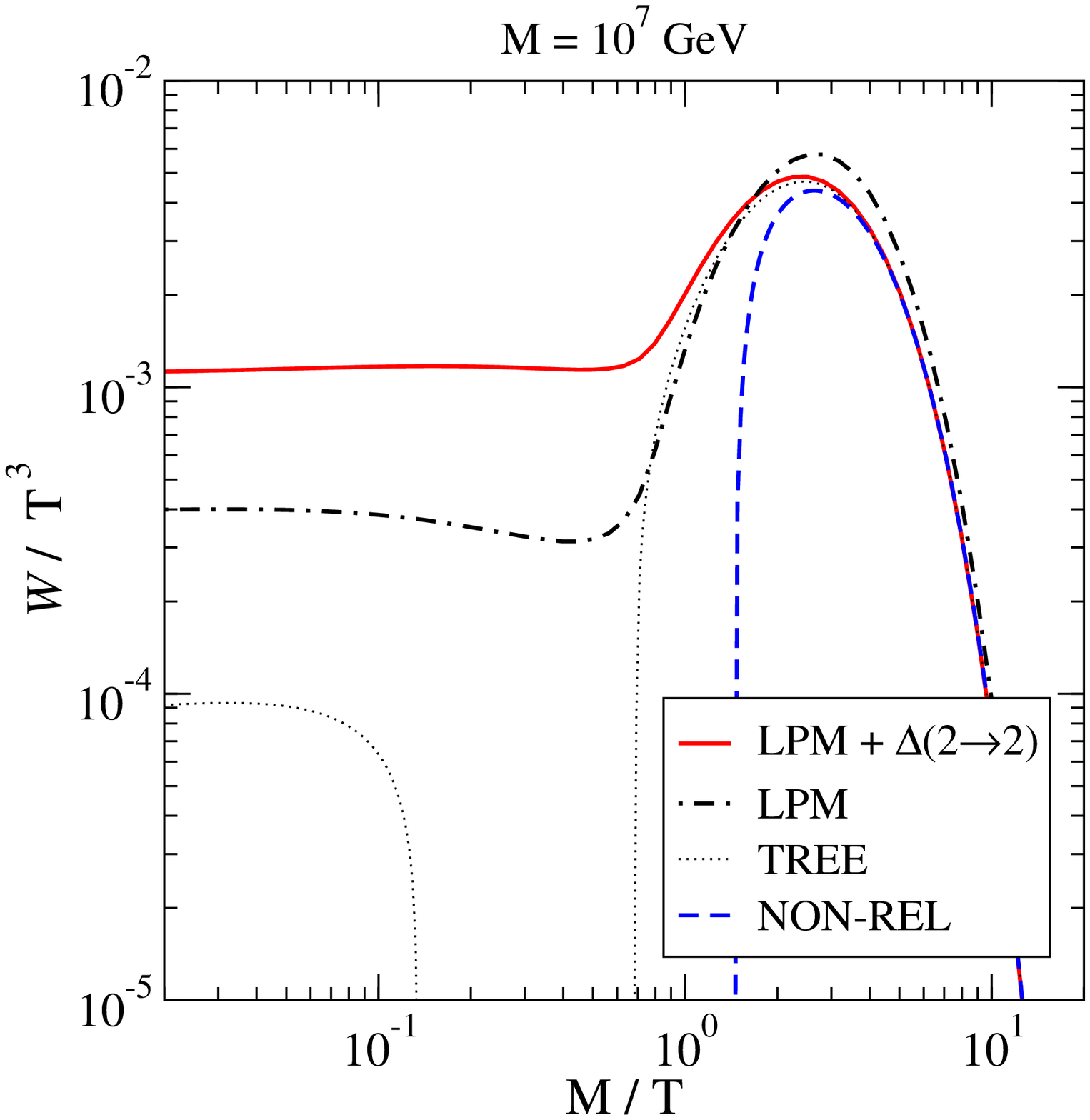}
}

\caption[a]{\small
Left: Total right-handed neutrino production rate from \eq\nr{gamma}, 
for $M = 10^7$~GeV. Shown are results from \eq\nr{final} 
(``LPM + $\Delta$\REAL'');
\eq\nr{lpm} (``LPM''); 
with naive thermal masses as given e.g.\ in \eq(3.9) 
of ref.~\cite{relat} (``TREE''); 
and from ref.~\cite{nonrel} (``NON-REL'').
Right: Similar results for the function defined in \eq\nr{washout}. 
% determining the lepton number washout rate
% as specified in ref.~\cite{washout}.
% The couplings and the renormalization scale 
% are fixed as specified in appendix~D.
The solid lines constitute our final results. 
}

\la{fig:rateRESUM}
\end{figure}
%%%%%%%%%%%%%%%%%%%%%%%%%%%%%%%%%%%%%%%%%%%%%%%%%%%%%%%%%%%%%%%%%%%%%%%%%%%

%%%%%%%%%%%%%%%%%%%%%%%%%%%%%%%%% FIGURE %%%%%%%%%%%%%%%%%%%%%%%%%%%%%%%%%
\begin{figure}[t]

%\vspace*{-3cm}

\centerline{%
\epsfysize=7.5cm\epsfbox{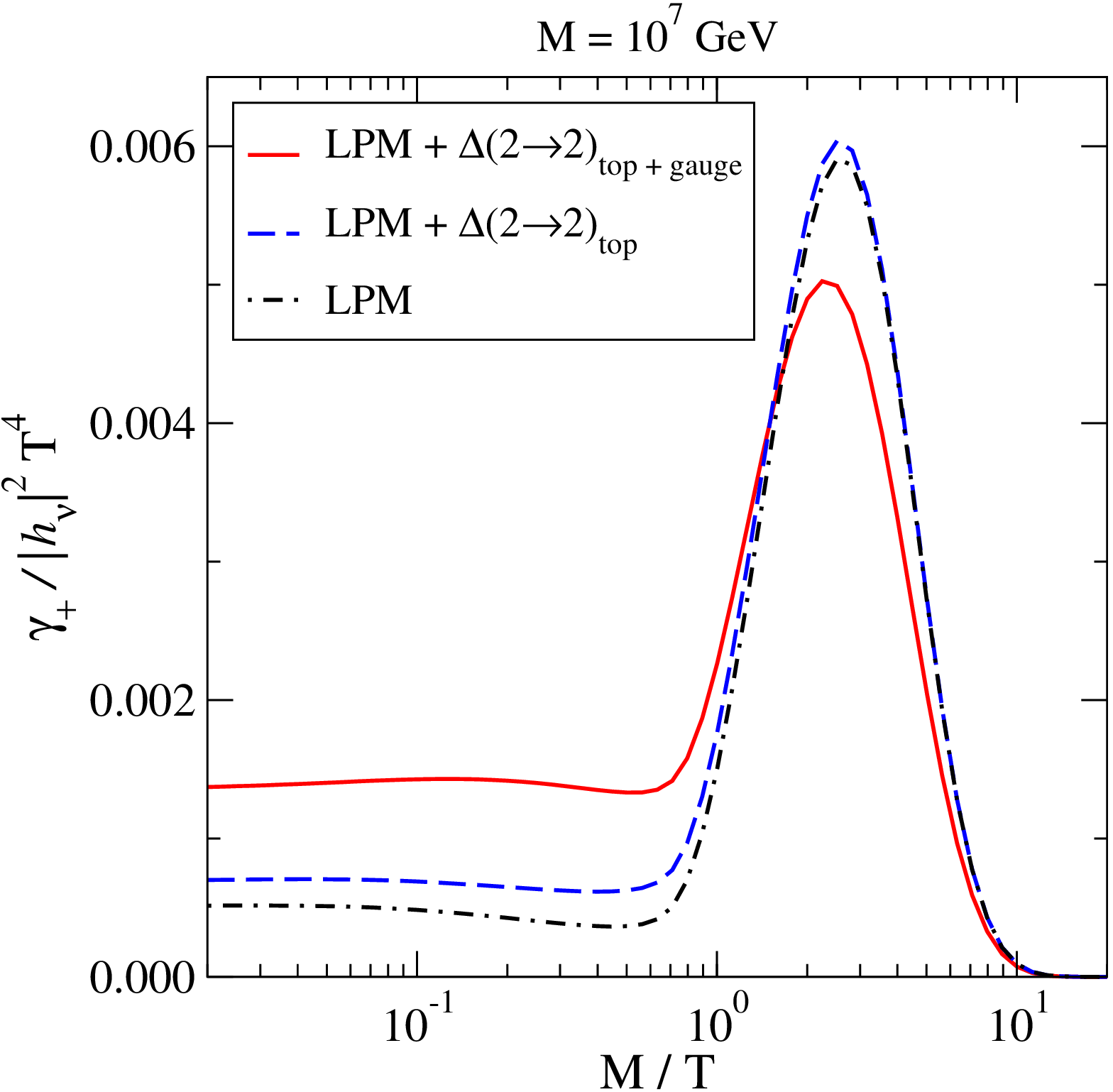}%
 ~~~\epsfysize=7.5cm\epsfbox{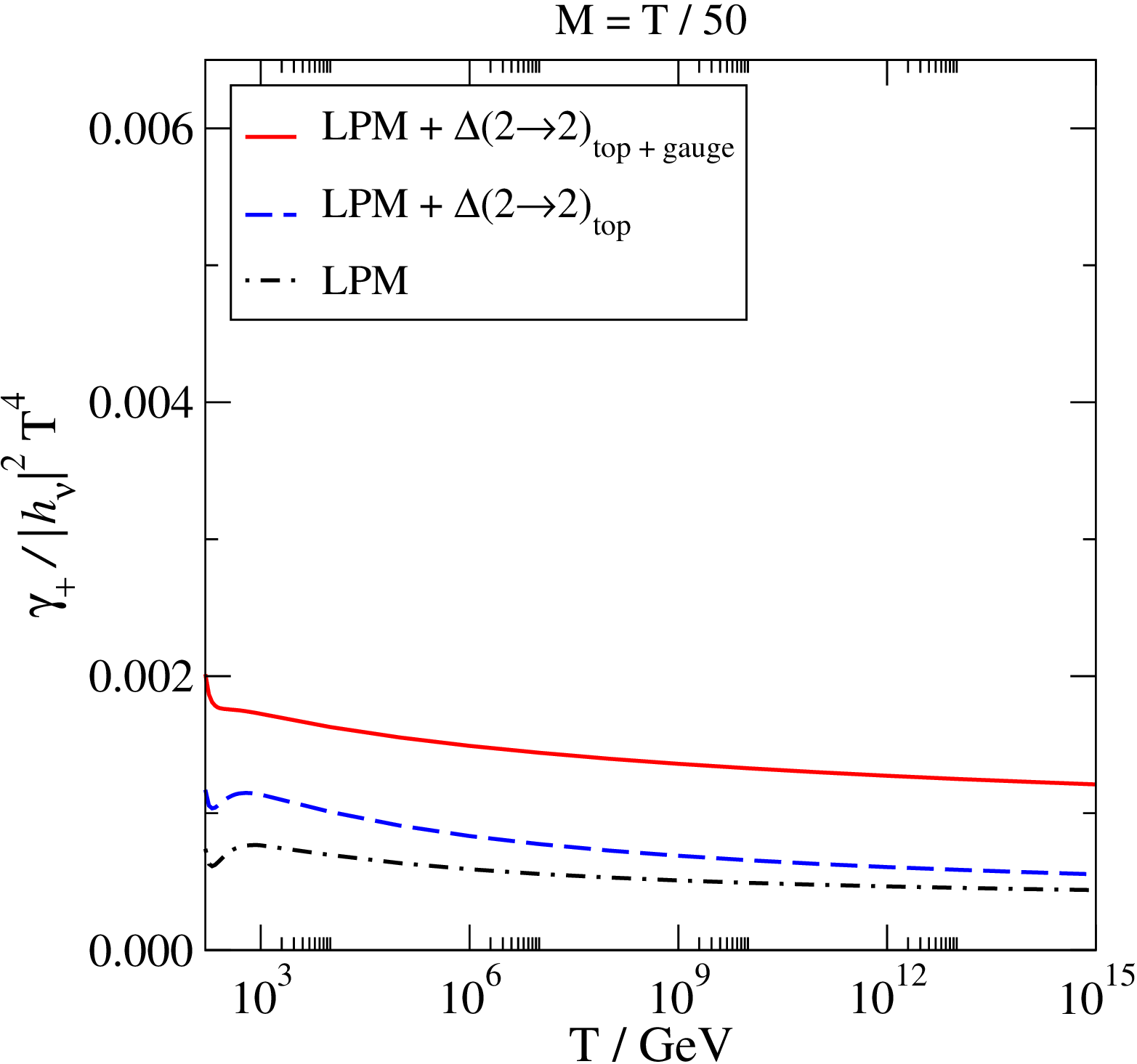}
}

\caption[a]{\small
Left: 
Like \fig\ref{fig:rateRESUM}(left) but with $\Delta$\REAL\ separated
into contributions from top-quark and gauge-boson scatterings. 
Right: 
Similar results but for fixed $M = 0.02T$, as a function of $T/$GeV.
}

\la{fig:parts}
\end{figure}
%%%%%%%%%%%%%%%%%%%%%%%%%%%%%%%%%%%%%%%%%%%%%%%%%%%%%%%%%%%%%%%%%%%%%%%%%%%

Let us proceed to numerical evaluations. 
In \fig\ref{fig:imRESUM}(left), the ``naive'' NLO and LPM 
results from \eqs\nr{nlo_phi}
and \nr{lpm} are shown; in \fig\ref{fig:imRESUM}(right), 
the difference defined by \eq\nr{delta_real}, which needs to 
be added to the LPM result, is displayed. The  
final estimate from \eq\nr{final} is shown in 
\fig\ref{fig:spectraRESUM}(left) as a function of $M/T$; and in 
\fig\ref{fig:spectraRESUM}(right) as a function of $k/T$.
In \figs\ref{fig:imRESUM}(right) and \ref{fig:spectraRESUM}(left), 
the uncertainty related to the phenomenological step in \eq\nr{nlo_phi}
is illustrated with a grey band (cf.\ the discussion around
\eq\nr{nlo_phi_tilde}). 
The total rates, from \eqs\nr{gamma}
and \nr{washout}, are shown in \fig\ref{fig:rateRESUM}.
Finally, in \fig\ref{fig:parts} we display on a linear
scale how the full result is made up of the LPM-resummed
$2\leftrightarrow 1$ result, and $2\to 2$ and $3\to 1$ scatterings 
as well as  
subleading corrections to $2\leftrightarrow 1$ processes
involving top quarks and gauge bosons. 

In addition to the figures, we have prepared a table containing
results with temperatures in the range 
$
 T = 
 (160 \ldots 10^{15}) \, \mbox{GeV} 
$,  
masses in the range 
$
 M =  (0.02 \ldots 20) \, T  
$, 
and momenta in the range 
$
 k =  (10^{-1} \ldots 10^{2} ) \, T
$.
The lowest temperature is determined by the location of 
the electroweak crossover in the Standard Model~\cite{Tc}.
The tabulated results can be downloaded as explained 
in footnote~\ref{fn:highT}. 

A few comments are in order: 
\begin{itemize}

\item
The physical spectra entering \eqs\nr{gamma}, \nr{washout}, {\it viz.} 
\ba
 \frac{ \partial_k\, \gamma_+}
 { |h_\nu|^2 }
 & \equiv &  
 \frac{k^2 \nF{}(\sqrt{k^2 +M^2})}{\pi^2 \sqrt{k^2 + M^2}} 
 \, \im \Pi^{ }_\rmii{R}
 \;, \la{dGamma} \\ 
 \partial_k\, W & \equiv & 
 - \frac{k^2 \nF{}'(\sqrt{k^2 +M^2})}{\pi^2 \sqrt{k^2 + M^2}} 
 \, \im \Pi^{ }_\rmii{R}
 \;, \la{dW}
\ea
can be obtained from the results displayed in 
\fig\ref{fig:spectraRESUM}(right) by trivial multiplications. 
% As is clearly seen from \fig\ref{fig:spectraRESUM}(right), 
% $\im \Pi^{ }_\rmii{R}$ depends
% only modestly on $k$ even in the ultrarelativistic regime $T \gg M$.

\item
Since empirically the results appear to have a smooth limit as $M/T \to 0$
(cf.\ \fig\ref{fig:spectraRESUM}(left)), expressions for the massless
case $M=0$ can to a good approximation be obtained from the row
of our table having $M = 0.02 T$. (Our numerics, optimized for 
a generic $M\sim T$, becomes ineffective for very small $M/T$, 
so we cannot set $M/T = 0$.)
Let us note, however, that for $M \lsim 0.1 T$
the spectrum shows an unphysical dip at $k \lsim 0.2 T$,
cf.\ \fig\ref{fig:spectraRESUM}(right). We believe this 
to be an artefact of the approximations made, which are 
based amongst others on the assumption that $k \gg g T$. Fortunately this
regime is phase-space suppressed in the observables
of \fig\ref{fig:rateRESUM}. If however spectral information plays an important 
role in a particular application, it is probably prudent to 
use the stable value $M \sim 0.2T$ 
rather than $M = 0.02 T$ as an approximation for $M/T = 0$.

\item
Typical numerical values of the thermal masses are 
$m_\phi \sim 0.4 T$ and $m^{ }_\ell \sim 0.3 T$. 
The rate at $M \ll T$ has a contribution from the 
$1\rightarrow 2$ decay $\phi\to \ell N$, since $m_\phi > 
M + m^{ }_\ell$ then. 
At temperatures close to the electroweak crossover, however, the
Higgs mass becomes small, cf.\ \eq\nr{mH}, and another channel
may open up. Therefore,  
the rate has a non-trivial shape at low temperatures, as is visible
in \fig\ref{fig:parts}(right). 

\item
For $M \ll T$ our results, as displayed in \figs\ref{fig:rateRESUM} 
and \ref{fig:parts}, are in good numerical agreement
with ref.~\cite{bb2}. In contrast, the results of ref.~\cite{bg2}
are larger by a factor $\sim 2$. It is difficult to identify a precise
reason for the discrepancy, but let us note that
if the considerations of \se\ref{se:X} were omitted, i.e.\ the 
doubly Bose-enhanced contributions from massless Higgs bosons 
were included in the NLO expression, our numerical results 
would be larger by a factor $\sim 1.7$ at $M \ll T$ (for $M = 10^7$~GeV).

\end{itemize}

%%%%%%%%%%%%%%%%%%%%%%%%%%%%% SECTION %%%%%%%%%%%%%%%%%%%%%%%%%%%%%%%%%%%%
%
\section{Conclusions and outlook}
\la{se:concl}

We have provided numerical results for the imaginary part of the right-handed
neutrino self-energy, entering gauge-invariant physical observables
as dictated by \eqs\nr{gamma} and \nr{washout}, as a function of 
the right-handed neutrino mass $M$ and momentum $k$, for a wide
range of temperatures $T \ge 160$~GeV.\footnote{%
 Tabulated results can be downloaded from \la{fn:highT}
 {www.laine.itp.unibe.ch/production-highT/}.
}  
Previous results for $M \ll T$~\cite{bb2} cannot be extrapolated to 
$M \gsim T$ because the $2\rightarrow 2$ contributions were evaluated
by assuming $M/T = 0$, whereas NLO 
results obtained for $M \gsim \pi T$~\cite{relat}
cannot be extrapolated to $M \ll \pi T$ because of a powerlike breakdown of 
the loop expansion.   
Our results smoothly interpolate between the two regimes, although
for the moment this comes with the price of a phenomenological treatment
in a particular intermediate range (cf.\ \se\ref{se:X}). 
In order to avoid this compromise in the future, the NLO computation
of ref.~\cite{relat} should be repeated with $m^{ }_\phi > 0$.  
{}From a practical point of view, though, it appears that only
a narrow mass range is affected, so that even the present results
should suffice for many applications (cf.\ the 
grey bands
in \figs\ref{fig:imRESUM}(right) and \ref{fig:spectraRESUM}(left), 
the latter being practically invisible). 

Apart from numerical evaluations, it would be highly desirable
to obtain analytic expressions as well. 
For the moment this has only been 
achieved as 
an expansion in a power series in $(\pi T/M)^2$, 
corresponding formally to an 
Operator Product Expansion~\cite{sch} and referred to as 
a non-relativistic regime~\cite{salvio,nonrel,tum}. 
Unfortunately, 
as discussed in ref.~\cite{relat}, this expansion shows poor convergence 
for $T \gsim M/15$, and is therefore not terribly useful for estimating  
thermal corrections in practice. 
Nevertheless it is helpful as 
a stringent crosscheck passed by the NLO 
expression~\cite{relat}, as well as a tool for 
formulating a theoretically consistent framework
for the full leptogenesis computation~\cite{bw}.

Let us end by noting that 
the imaginary part of the right-handed neutrino 
self-energy has also been studied at temperatures 
below about 10~GeV~\cite{numsm}.\footnote{%
 Tabulated results can be downloaded from \la{fn:lowT}
 {www.laine.itp.unibe.ch/production-lowT/}.
}   
It is a relevant challenge for
future work to close the gap between $T\lsim 10$~GeV
and $T\gsim 160$~GeV.

%%%%%%%%%%%%%%%%%%%%%%%%% SECTION %%%%%%%%%%%%%%%%%%%%%%%%%%%%%%%%%%%%%
%
\section*{Acknowledgements}

M.L is grateful to D.~B\"odeker for helpful discussions. 
This work was partly supported by the Swiss National Science Foundation
(SNF) under grant 200020-155935.

%%%%%%%%%%%%%%%%%%%%%%% APPENDIX %%%%%%%%%%%%%%%%%%%%%%%%%%%%%%%%%%%
%
\appendix
\renewcommand{\thesection}{Appendix~\Alph{section}}
\renewcommand{\thesubsection}{\Alph{section}.\arabic{subsection}}
\renewcommand{\theequation}{\Alph{section}.\arabic{equation}}

%%%%%%%%%%%%%%%%%%%%%%%%%%%%%% SECTION %%%%%%%%%%%%%%%%%%%%%%%%%%%%%%%%%
%
\section{Real corrections within the NLO expression}
\la{app:A}

In order to allow for a comparison with Boltzmann equations, we re-write
here the ``real corrections'' appearing in \eq\nr{nlo} in the form of matrix
elements squared. The normalization is chosen so as to permit for a direct
comparison with the expressions given in table~1 of ref.~\cite{bb2}.
(Let 
us stress again that the $2\to 2$ and $3\to 1$ corrections are not integrable
as such, but that the expression as a whole is finite, provided that
its 1-loop corrected $2\to 1$  part is added in the presence of 
a consistent regulator for soft momentum transfer.) 

By making use of the results listed in refs.~\cite{relat,master}, 
\eq\nr{nlo} can be re-expressed as 
\ba
 && \hspace*{-1cm} 2 \nF{}(\ko) \Bigl[ 
   \im \Pi^\rmii{NLO}_\rmii{R} - 
   \im \Pi^\rmii{LO}_\rmii{R}
 \Bigr]
 \nn 
 & = & 
 \int \! {\rm d}\Omega^{ }_{2\to 1} \;
% \nF{}(p_1) \, \nB{}(p_2) \, 
 \Bigl\{
  \mbox{\eq\nr{virtual}} 
 \Bigr\}
 \nn 
 & + & 
 \int \! {\rm d}\Omega^{ }_{2\to 2} \; 
 \Bigl\{\; 
  \nB{}(p_1) \, \nB{}(p_2) \, \bigl[1 - \nF{}(k_2)\bigr]
  \, |\mathcal{M}^{ }_\rmi{a}|^2 
 \nn & & \hspace*{1.6cm} + \, 
   \nF{}(p_1) \, \nB{}(p_2) \, \bigl[1 + \nB{}(k_2)\bigr]
  \, \sum |\mathcal{M}^{ }_\rmi{b}|^2 
 \nn & & \hspace*{1.6cm} + \, 
   \nF{}(p_1) \, \nF{}(p_2) \, \bigl[1 - \nF{}(k_2)\bigr]
  \, \sum |\mathcal{M}^{ }_\rmi{c}|^2 
 \; \Bigr\} 
 \nn 
 & + & 
 \int \! {\rm d}\Omega^{ }_{3\to 1} \; 
 \Bigl\{\; 
  \nF{}(p_1) \, \nB{}(p_2) \, \nB{}(p_3) 
  \, |\mathcal{M}^{ }_\rmi{d}|^2 
 \nn & & \hspace*{1.6cm} + \, 
   \nF{}(p_1) \, \nF{}(p_2) \, \nF{}(p_3)
  \, |\mathcal{M}^{ }_\rmi{e}|^2 
 \; \Bigr\} 
 \;. \la{boltzmann}
\ea
Here ${\rm d}\Omega^{ }_{n\to m}$ denotes the usual phase space 
integration measure with 4-momentum conservation, 
${\rm d}\Omega^{ }_{n\to m} \equiv 
 \Pi_{i=1}^{n} \frac{{\rm d}^3\vec{p}_i}{2 p_i (2\pi)^3 }
 \Pi_{j=2}^{m} \frac{{\rm d}^3\vec{k}_j}{2 k_j (2\pi)^3 }
 \, (2\pi)^4 \, \delta^{(4)}  
 ( \sum_{i=1}^n \mathcal{P}_i - \sum_{j=1}^m \mathcal{K}_j
 )
$. 
The three-momenta of incoming particles are denoted by $\vec{p}_i$, 
with $p_i \equiv |\vec{p}_i|$ and 
$\mathcal{P}^{ }_i \equiv (p_i,\vec{p}_i)$; those of outgoing particles are 
$\vec{k}_i$, with $\vec{k}_1 \equiv \vec{k}$ the right-handed
neutrino momentum. The matrix elements squared read
\ba
 |\mathcal{M}^{ }_\rmi{a}|^2 & \equiv & 
 \bigl(g_1^2 + 3 g_2^2\bigr) 
 \, \biggl(  \frac{u - M^2}{t} - \frac{2 u M^2}{st} \biggr)
 \;, \\ 
 \sum |\mathcal{M}^{ }_\rmi{b}|^2 & \equiv & 
 \bigl(g_1^2 + 3 g_2^2\bigr) 
 \, \biggl(
  \frac{M^2 - u}{s} 
+ \frac{M^2 - s}{u} 
+ \frac{2 u M^2}{st}
+ \frac{2 s M^2}{ut}
 \biggr)
 \;, \\ 
 \sum |\mathcal{M}^{ }_\rmi{c}|^2 & \equiv & 
 2 h_t^2 \Nc
 \, \biggl(
  3 - \frac{M^2}{s} - \frac{2 M^2}{t} 
 \biggr)
 \;, \\ 
 |\mathcal{M}^{ }_\rmi{d}|^2 & \equiv & 
 \bigl(g_1^2 + 3 g_2^2\bigr) \, 
 \biggl( 
 \frac{M^2 - s_{13}}{s_{12}} 
+ \frac{2 s_{13}  M^2}
 {s_{12} s_{23} }
 \biggr) \;, \\ 
 |\mathcal{M}^{ }_\rmi{e}|^2 & \equiv & 
 2 h_t^2 \Nc \, 
 \biggl( 
 -1 + \frac{M^2}{ s_{23} } 
 \biggr)
 \;. \la{M*M}
\ea 
Here 
$s \equiv (\mathcal{P}_1 + \mathcal{P}_2)^2$,
$t \equiv (\mathcal{K}_2 - \mathcal{P}_2)^2$,
$u \equiv (\mathcal{K}_2 - \mathcal{P}_1)^2$,
and 
$s_{ij} \equiv (\mathcal{P}_i + \mathcal{P}_j)^2$; 
these quantities are related through
$ 
 s + t + u = s_{12} + s_{13} + s_{23} = M^2
$. 
Setting $M\to 0$ the $2\to 2$ matrix elements 
agree with those in ref.~\cite{bb2}, 
with $u\leftrightarrow t$ in the $s$-channel case.

%%%%%%%%%%%%%%%%%%%%%%%%%%%%%% SECTION %%%%%%%%%%%%%%%%%%%%%%%%%%%%%%%%%
%
\section{Virtual corrections within the NLO expression}
\la{app:B}

We define a $2\rightarrow 1$ integration measure like in appendix~A, 
but in the presence of dimensional regularization and after the 
introduction of an infrared regulator $\lambda$ into the lepton energy:
\be
 {\rm d}\Omega^{ }_{2\rightarrow 1} \equiv 
 \frac{{\rm d}^{3-2\epsilon} \vec{p}_1}{2 \epsilon_1 (2\pi)^{3-2\epsilon}}
 \, \frac{{\rm d}^{3-2\epsilon} \vec{p}_2}{2 p_2 (2\pi)^{3-2\epsilon}}
 \, (2\pi)^{4-2\epsilon} (\mathcal{P}_1 + \mathcal{P}_2 - \mathcal{K})
 \;, 
\ee
where $\epsilon_1 \equiv \sqrt{p_1^2 + \lambda^2}$ and 
$\mathcal{P}_1 \equiv (\epsilon_1,\vec{p}_1)$. Then the $2\rightarrow 1$
part of the NLO result in \eq\nr{nlo} can formally be written as 
(for $\ko > k$)
\ba
 & & \hspace*{-1cm}
  \Bigl[ \im \Pi^\rmii{NLO}_\rmii{R} - 
   \im \Pi^\rmii{LO}_\rmii{R} \Bigr]^{ }_{2\to 1}
 \; = \; 
 \lim_{\lambda\to 0}
 \int \! {\rm d}\Omega^{ }_{2\to 1} \;
 \Bigl[ 1 - \nF{}(\epsilon_1) + \nB{}(p_2) \Bigr] \, 
 \la{virtual} \\[3mm]
% & & \; \times \, 
  & \times & 
 \mathbbm{P} \, 
 \biggl\{
 \, m_\ell^2 \, \biggl[ 1 + M^2
  \frac{ \overleftarrow{ {\rm d} } }{{\rm d}\lambda^2}  \biggr]
%
% \nn & &
 \; 
  -  \, h_t^2 \Nc 
    \Tint{\{ Q \}} \frac{ M^2  }{Q^2(Q-P^{ }_2)^2}
    \, \biggr|^{ }_{p^{ }_{2n} = - i p_2}
 \nn  
%  & & \; + \, 
  & + & 
  (g_1^2 + 3 g_2^2)  
  \biggl[
    - \Tint{ Q } \frac{M^2}{Q^2(Q-K)^2}
    \, \biggr|^{ }_{k_{n} = - i \ko }
    + \Tint{ Q } \frac{M^2}{Q^2(Q-P^{ }_2)^2}
    \, \biggr|^{ }_{p^{ }_{2n} = - i p_2}
 \nn
% & & \;  + \,
  & + & 
 \Tint{ Q } \frac{M^2 + (1-\epsilon)\, Q\cdot K }{Q^2(Q-P^{ }_1)^2}
    \, \biggr|^{ }_{p^{ }_{1n} = - i p_1}
%
%    + \Tint{ Q } \frac{(1-\epsilon)\, Q\cdot K}{Q^2(Q-P^{ }_1)^2}
%    \, \biggr|^{ }_{p^{ }_{1n} = - i p_1}
%    
% \nn
%  & & \;  + \,
% & + & 
 + 
 \Tint{ Q } \frac{M^4}{Q^2(Q-P^{ }_2)^2(Q-K)^2}
    \, \biggr|^{ }_{p^{ }_{2n} = - i p_2,\, k_n = - i \ko}
 \;
 \biggr]
 \biggr\}
% \nn 
% & + & 
% \int \! {\rm d}\Omega^{ }_{2\to 2} \; 
% \Bigl\{\; \ldots  \; \Bigr\} 
%  + 
% \int \! {\rm d}\Omega^{ }_{3\to 1} \; 
% \Bigl\{\; \ldots  
% \; \Bigr\} 
 \;, \nonumber 
\ea
where $\mathbbm{P}$ refers to a principal value, and 
$\{Q\}$, $P^{ }_1$ and $K$ are fermionic Matsubara four-momenta
(with $P^{ }_i \equiv (p^{ }_{in},\vec{p}^{ }_i)$ etc).
The thermal lepton mass is given by \eq\nr{mH}, and 
can also be expressed as 
\be
 m_\ell^2 = \frac{g_1^2 + 3 g_2^2}{2}
 \int_\vec{q} \frac{\nB{}(q) + \nF{}(q)}{q}
 \;. \la{mell}
\ee
With a similar notation, the LO part can be expressed as 
(for $m^{ }_\phi\to 0$) 
\be
 \lim_{m^{ }_\phi\to 0} \im \Pi_\rmii{R}^\rmii{LO} = M^2 \lim_{\lambda \to 0}
 \int \! {\rm d}\Omega^{ }_{2\to 1} 
 \, 
 \Bigl[ 1 - \nF{}(\epsilon_1) + \nB{}(p_2) \Bigr] 
 \;.  \la{lo_naive}
\ee

Comparing \eq\nr{virtual} with \eq\nr{lo_naive}, 
most of the terms in \eq\nr{virtual} would appear to be NLO corrections 
to the LO result. Indeed, evaluating the sum-integrals in the limit 
of zero temperature and keeping only the $1/\epsilon$-divergences, 
it can readily be checked that divergences are those 
cancelled by $\mathcal{Z}^{ }_\nu$ of \eq\nr{Z_nu}. However, 
because momenta are set on-shell in the structures
appearing in \eq\nr{virtual}, 
the results are infrared divergent (even at zero temperature). 
The full 
result is finite only once summed together with the real processes. 

There are a few structures in \eq\nr{virtual} which are 
particularly important at high temperatures, being of relative 
magnitude $\sim g^2 T^2/M^2$ in this limit. 
These are the so-called Hard Thermal Loops (HTLs). Apart from the terms
already expressed as $m_\ell^2$, the only other HTL originates from  
$\Tinti{Q} \frac{ Q\cdot K }{Q^2(Q-P_1)^2}$~\cite{brpi}.\footnote{% 
 Note that terms leading to a thermal Higgs mass were already 
 taken away into \eq\nr{tree}.
 }
It turns out, however, that in reality even some non-HTL structures
lead to a similarly large final result, 
despite the apparent prefactor $M^2$. This issue
is discussed in \se\ref{se:X}. Here we show  
that the ``normal'' HTL structures in \eq\nr{virtual} 
amount exactly to the HTL resummation of the lepton propagator. 

The HTL-resummed inverse lepton propagator reads~\cite{kli,haw}
$
 \Sigma(P) = i \, \bigl[ 
  \bsl{P}  + \frac{m_\ell^2}{2}
  \int_\vec{v} \frac{i \gamma_0 + \vec{v}\cdot\bm{\gamma}}
                    {i p_n + \vec{v} \cdot \vec{p}}
 \bigr]
$, 
where $|\vec{v}| = 1$,  $\int_\vec{v} 1 = 1$, and 
we employ Euclidean Dirac-matrices. If the HTL self-energy is treated
as an insertion, the lepton propagator reads
\be
 \Sigma^{-1}(P) = 
 - \frac{i \bsl{P}}{P^2} + 
 \frac{i m_\ell^2}{2 }
 \frac{\bsl{P}}{P^2}
  \int_\vec{v} \frac{i \gamma_0 + \vec{v}\cdot\bm{\gamma}}
                    {i p_n + \vec{v} \cdot \vec{p}}
 \,\frac{\bsl{P}}{P^2}
 \;. 
\ee
Determining the correlator of \eq\nr{Pi} with this propagator, we find
\be
 \Pi_\rmii{E}^\rmii{HTL,ins} 
 \equiv \lim_{\lambda\to 0}
 \int_{\vec{v}}\Tint{P}
 \frac{2}{(P-K)^2}
 \biggl\{
  -\biggl( 1 + m_\ell^2 
  \frac{{\rm d}}{{\rm d}\lambda^2} \biggr) \frac{2 K\cdot P}{P^2 + \lambda^2}  
  -  
   \frac{m_\ell^2\, (i k_n + \vec{v}\cdot\vec{k} )}
   {(P^2+\lambda^2)(i p_n + \vec{v}\cdot\vec{p}) }  
 \biggr\}
 \;, \la{Pi_HTL}
\ee
where $\lambda$ has been introduced as an infrared regulator. 

In the last term of \eq\nr{Pi_HTL}, there are two 
``inverse propagators'', 
$
  P^2+\lambda^2 
$ 
and 
$ 
 i p_n + \vec{v}\cdot\vec{p}
$. 
If the 
part $P^2+\lambda^2$ is cut, we get ``pole'' contributions, whereas
cutting $i p_n + \vec{v}\cdot\vec{p}$ yields a ``cut'' contribution. 
These can be separated by partial fractioning
($
 \epsilon^{ }_1 \equiv \sqrt{p^2 + \lambda^2}
$), 
\ba
 \frac{1}{(P^2+\lambda^2)(ip_n + \vec{v}\cdot\vec{p})} 
 & = & \frac{1}{2\epsilon^{ }_1}
 \biggl[
   \frac{1}{(\epsilon^{ }_1 + \vec{v}\cdot\vec{p})(\epsilon^{ }_1 - i p_n)} 
 - 
   \frac{1}{(\epsilon^{ }_1 - \vec{v}\cdot\vec{p})(\epsilon^{ }_1 + i p_n)} 
 \biggr]
 \nn 
 & + & 
   \frac{1}{[\epsilon_1^2 - (\vec{v}\cdot\vec{p})^2]
   (i p_n + \vec{v}\cdot\vec{p})} 
 \;.  \la{fraction}
\ea
The last term represents the soft momentum transfer regime of 
$2\leftrightarrow 2$ scatterings, and 
is analyzed in more detail in appendix~C. 
The structure 
$1/(\epsilon^{ }_1 - i p_n)$ leads 
to $\delta(\ko - |\vec{p}-\vec{k}| + \epsilon^{ }_1)$
after carrying out the sum over $p_n$ and taking the cut, 
which does not get realized for $\ko > k$.
Therefore the only $2\leftrightarrow 1$ correction arises from the second 
term of \eq\nr{fraction}. Summing this together with  
the first term in \eq\nr{Pi_HTL} and noting that
\be
 \lim_{\lambda\to 0}
 \biggl( 1 + m_\ell^2 
  \frac{{\rm d}}{{\rm d}\lambda^2} \biggr) \, (M^2 + \lambda^2 )
 \, f(\lambda^2)
 = M^2 f(0) +  \lim_{\lambda\to 0} m_\ell^2 
 \biggl( 1 + M^2 
  \frac{{\rm d}}{{\rm d}\lambda^2} \biggr) 
 \, f(\lambda^2)
 \;, 
\ee
the $2\leftrightarrow 1$ part of \eq\nr{Pi_HTL} can be re-written as
($
 \epsilon^{ }_1 \equiv \sqrt{p^2 + \lambda^2}
$, 
$
 \epsilon^{ }_2 \equiv |\vec{p}-\vec{k}|
$)
\ba
 \im \Pi_\rmii{R}^\rmii{HTL,ins,pole} & \equiv & 
 \lim_{\lambda\to 0} \int_{\vec{v},\vec{p}}
 \biggl\{ 
  \;
   M^2 + m_\ell^2  
  \biggl( 1 + M^2 
  \frac{{\rm d}}{{\rm d}\lambda^2}
   -  \; \frac{\ko + \vec{v}\cdot\vec{k}}
  {\epsilon^{ }_1 + \vec{v}\cdot\vec{p}}
   \biggr)  \;
 \biggr\}
 \nn & \times & 
  \frac{2\pi \delta(\ko - \epsilon^{ }_1 - \epsilon^{ }_2)}
  {4 \epsilon^{ }_1 \epsilon^{ }_2 }
  \, 
  \Bigl[ 1 - \nF{}(\epsilon^{ }_1) + \nB{}(\epsilon^{ }_2)\Bigr]
 \;. \la{HTLpole}
\ea

It is now easy to check that the $m_\ell^2$-parts of \eq\nr{HTLpole}
agree with the HTL parts of \eq\nr{virtual}, once the structure
$\Tinti{Q} \frac{ Q\cdot K }{Q^2(Q-P_1)^2} |^{ }_{p^{ }_{1n} = - i p_1}$ 
is expanded to leading
order in $p_1/T$; the thermal lepton mass is 
identified as \eq\nr{mell}; and a velocity variable is 
defined as $\vec{v} \equiv \vec{q}/q$.

%%%%%%%%%%%%%%%%%%%%%%%%%%%%%% SECTION %%%%%%%%%%%%%%%%%%%%%%%%%%%%%%%%%
%
\section{Region of soft momentum transfer}
\la{app:C}

At zero temperature it is well established that infrared (collinear and 
soft) divergences associated with real and virtual corrections cancel 
in physical observables. As was observed in refs.~\cite{relat,master} 
(generalizing on a previous analysis at vanishing
spatial momentum~\cite{bulk_wdep}), a similar cancellation 
takes place in every ``master'' spectral function 
at finite temperature. As a part of the analysis
of \se\ref{se:subtraction}, we have shown that the cancellation also
takes place in the ultrarelativistic regime, where it can be treated
within the HTL effective theory (specifically, this refers to the
cancellation of $\lambda$ in \eq\nr{virtual_2}). 
In this appendix we provide 
more details concerning the HTL setup, and also compute the ``cut'' 
contributions needed in \se\ref{se:subtraction}, representing the
effects of $2\to 2$ scatterings mediated by 
soft $t$-channel leptons. 

If we consider the correlator of \eqs\nr{Pi}, \nr{relation} within 
the HTL theory, it may first be verified (by explicit computation) 
that there is no vertex correction. In the Higgs propagator the 
only change is the appearance of a thermal mass.  Denoting by 
$\rho^{ }_\ell(\omega,\vec{p})$ the spectral function corresponding
to the lepton propagator, we find
\be
 \im \Pi^\rmii{HTL,full}_\rmii{R}  \equiv  
 \int_{-\infty}^{\infty}\!\!\!\! {\rm d}\omega \! \int_{\vec{p}}
 \frac{- 2 \mathcal{K}\cdot \rho^{ }_\ell(\omega,\vec{p})}
 {\epsilon^{ }_2} \, 
 \Bigl[ 1 - \nF{} (\omega) + \nB{}(\epsilon^{ }_2) \Bigr]
 \delta\bigl( \ko - \omega - \epsilon^{ }_2 \bigr)
 \;, \la{HTL}
\ee
where only that pole from the Higgs propagator which gets realized 
in practice has been kept, and 
$\epsilon^{ }_2 \equiv \sqrt{(\vec{p-k})^2 + m_\phi^2} $. 
The spectral function (defined as a four-vector) can be expressed as 
\be
 \rho^{ }_\ell (\omega,\vec{p}) \equiv
 \Bigl( \omega\, \hat{\rho}_0 (\omega,p)
 ,\vec{p}\, \hat{\rho}_s (\omega,p)
 \, \Bigr)
 \;. 
\ee
Carrying out the angular integral in \eq\nr{HTL} yields 
\ba
 \im \Pi^\rmii{HTL,full}_\rmii{R} & = & -\frac{1}{2\pi^2 k}
 \int_{-\infty}^{\infty}\!\! {\rm d}\omega \! 
  \int_{p_\rmii{min}(\omega)}^{p_\rmii{max}(\omega)} \! {\rm d}p \, p \, 
 \Bigl[ 1 - \nF{} (\omega) + \nB{}(\ko - \omega) \Bigr]
 \nn & \times & 
 \biggl\{ \ko\, \omega \bigl[ \hat{\rho}^{ }_0 - \hat{\rho}^{ }_s \bigr]
  + \frac{M^2 - m_\phi^2 + \omega^2 - p^2}{2} \, \hat{\rho}^{ }_s \biggr\}
% \theta(\ko - \omega - \sqrt{(k-q)^2 + m_\phi^2})
 \;. \la{HTL2}
\ea

The spectral functions $\hat{\rho}^{ }_0, \hat{\rho}^{ }_s$ have a 
``pole part'' at $|\omega| > p$ and a ``cut part'' at 
$|\omega| < p$. The latter is often referred to as Landau damping, 
and reflects the effects of $1\leftrightarrow 2$ scatterings 
of off-shell leptons on thermal gauge bosons, which are a subpart
of $2\rightarrow 2$ scatterings in \fig\ref{fig:processes}. 
The explicit forms of the spectral functions read
\ba
 \hat{\rho}^{ }_0 & = & 
 \im \Biggl\{ \frac{1 - \frac{m_\ell^2 L}{2\omega}}
 {\bigl[\omega - \frac{ m_\ell^2 L}{2}\bigr]^2 -
  \bigl[p + \frac{m_\ell^2 (1-\omega L)}{2 p}\bigr]^2} \Biggr\}
 \;, \la{rho0} \\
 \hat{\rho}^{ }_s & = & 
 \im \Biggl\{ \frac{ 1 + \frac{m_\ell^2 (1-\omega L)}{2 p^2}  }
 {\bigl[\omega - \frac{ m_\ell^2 L}{2}\bigr]^2 -
  \bigl[p + \frac{m_\ell^2 (1-\omega L)}{2 p}\bigr]^2} \Biggr\}
 \;, \la{rhos}
\ea
where $L \equiv \frac{1}{2p} \ln \frac{\omega + p}{\omega - p}$ 
and $\omega$ has a small positive imaginary part.
The corresponding phase space is illustrated in \fig\ref{fig:phasespace}.
(There are actually two poles, although
only one contributes in the regime $p \gg m^{ }_\ell$.
Both poles, but no cuts, were included in ref.~\cite{mht}.) 

%%%%%%%%%%%%%%%%%%%%%%%%%%%%%%%%% FIGURE %%%%%%%%%%%%%%%%%%%%%%%%%%%%%%%%%
\begin{figure}[t]

%\vspace*{-3cm}

\centerline{%
 \epsfysize=6.5cm\epsfbox{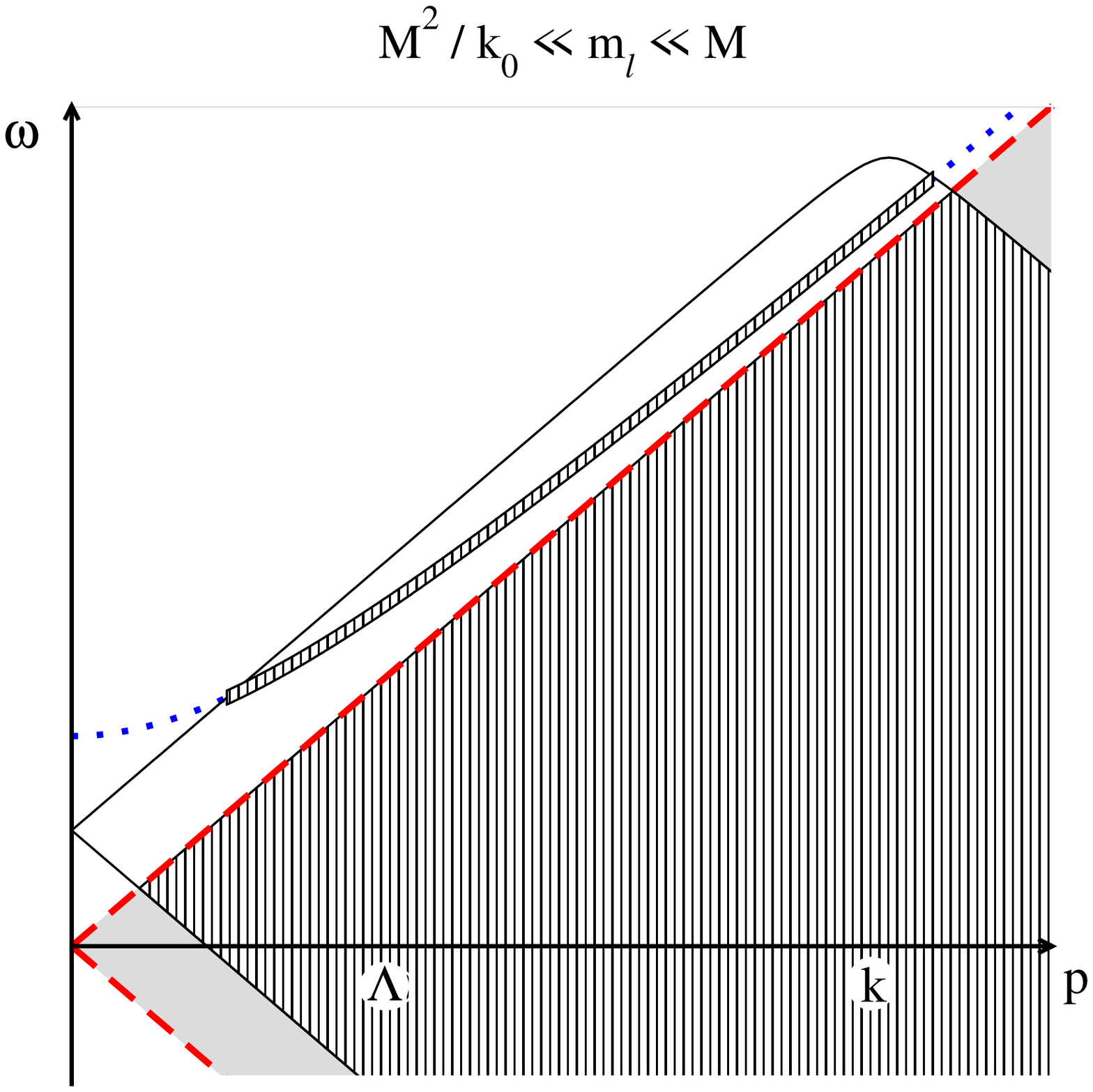}%
~~~~~~~\epsfysize=6.5cm\epsfbox{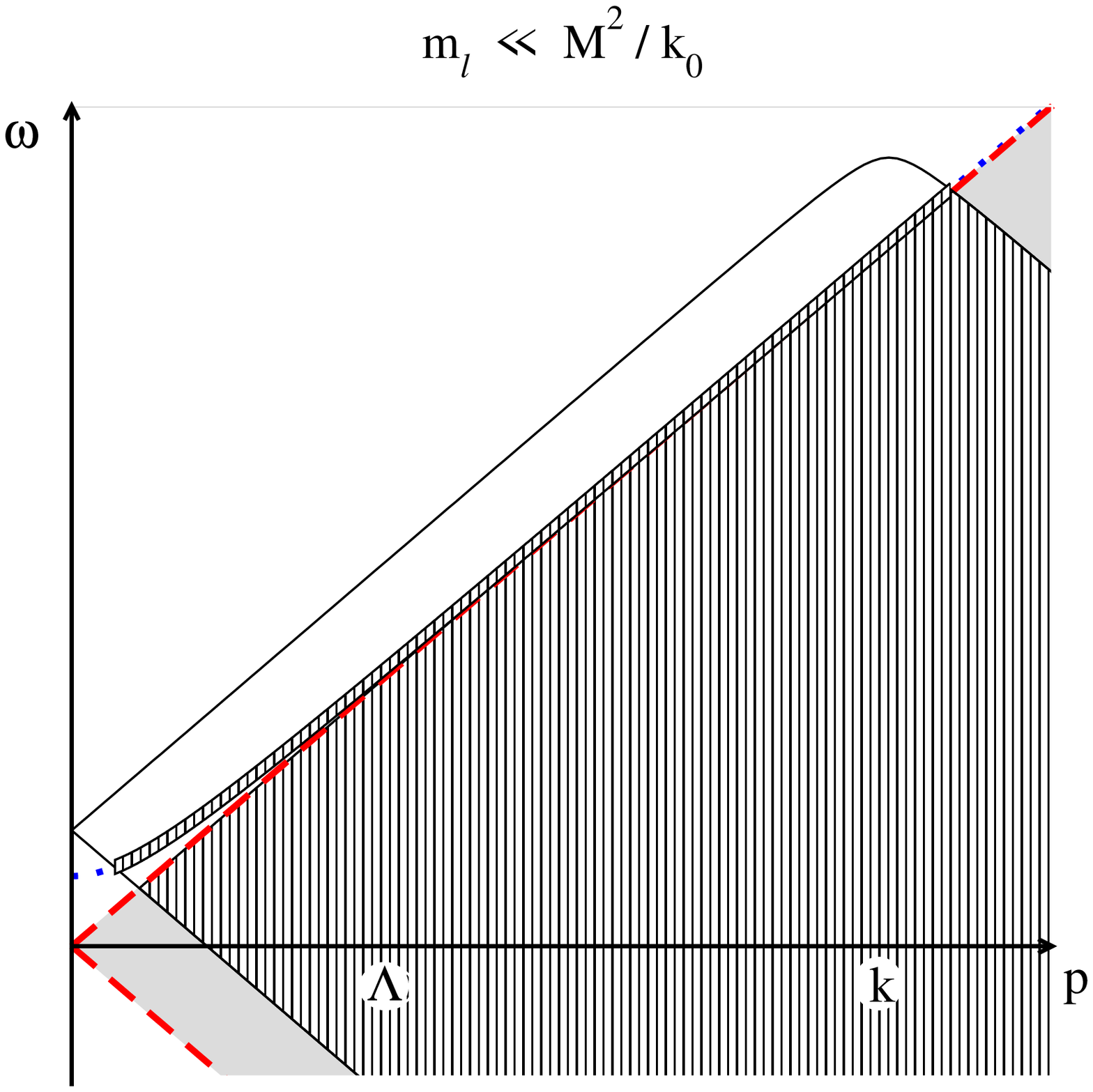}
}

\caption[a]{\small
An illustration of the phase space relevant for HTL resummation. 
% It is assumed that $M > m_\phi + m^{ }_\ell$. 
The grey-shaded area and the dotted
blue line indicate regions in which $\rho^{ }_\ell$ 
is non-zero (only one pole is shown). The solid black line, 
with intercept $\omega = \ko - \sqrt{k^2 + m_\phi^2}$ at $p=0$,  
delineates the region allowed
kinematically by the $\delta$-constraint in \eq\nr{HTL}. The hashed
areas denote regions contributing to \eq\nr{HTL2}. 
% On the left, the case $M^2/\ko \ll m^{ }_\ell \ll M$ is shown, 
% whereas on the right, $m^{ }_\ell \ll M^2/\ko$.
}

\la{fig:phasespace}
\end{figure}
%%%%%%%%%%%%%%%%%%%%%%%%%%%%%%%%%%%%%%%%%%%%%%%%%%%%%%%%%%%%%%%%%%%%%%%%%%%

Let us now consider the contribution of soft momenta ($p \lsim gT$) to 
$ \im \Pi^\rmii{HTL}_\rmii{R} $. For this we introduce a cutoff
$\Lambda$, satisfying 
$m^{ }_\ell \ll \Lambda  \ll k$
(cf.\ \fig\ref{fig:phasespace}). In the soft region we can also 
set $m_\phi = 0$, because the Higgs momentum $\vec{k}-\vec{p}$
is hard ($p \ll k$); in fact, for the analysis of 
\se\ref{se:subtraction} we {\em need}
to set $m_\phi = 0$, because this is the case
in \eq\nr{nlo}.  
We work within the 
kinematics relevant for the ultrarelativistic regime, i.e.\ $M \ll k$, 
so that $\ko \approx k$.

In this appendix we focus on soft $2\to 2$ scatterings which, 
as mentioned,  
correspond to the {\em cut} contribution in the HTL setup. 
We consider the cut contribution in two different ways. 
First we consider it in the ``full'' form as dictated
by the HTL theory. Second, we consider it in the ``inserted'' form in which it 
appears in the $2\to 2$ part of the 
NLO result of \se\ref{se:nlo}, where the scale $m^{2}_\ell$
only appears as an overall prefactor. 

%%%%%%%%%%%%%%%%%%

For $|\omega| < p$, the function $L$
has an imaginary part, $\im L = -\pi / (2 p)$. The cut originates from 
this imaginary part, and is necessarily proportional
to $m_\ell^2$. Therefore, omitting higher-order corrections, 
we can put $M\to 0$ in $p^{ }_\rmi{min} = \km \approx M^2/(4\ko)$. 
Furthermore, the second term in \eq\nr{HTL2} is subleading, 
given that $\hat{\rho}^{ }_s$ is antisymmetric in $\omega$.
Therefore the first term gives the leading contribution: 
\ba
 \im \Pi^\rmii{HTL,full,cut}_\rmii{R} \!\! & \equiv & \!\! 
 -\frac{1}{2\pi^2} \int_0^\Lambda \! {\rm d}p \, p 
 \int_{-p}^p \! {\rm d}\omega \, \omega \, 
 \Bigl[ \nB{}(\ko) + \fr12 \Bigr]
 \im \Biggl\{ \frac{ \frac{m_\ell^2 (\omega^2 - p^2) L}{2 \omega p^2} - 
 \frac{m_\ell^2}{2 p^2}  }
 {\bigl[\omega - \frac{ m_\ell^2 L}{2}\bigr]^2 -
  \bigl[p + \frac{m_\ell^2 (1-\omega L)}{2 p}\bigr]^2} \Biggr\}
 \nn 
 & = & \frac{m_\ell^2}{4\pi} 
 \Bigl[ \nB{}(\ko) + \fr12 \Bigr]
 \biggl[ 
  \ln\Bigl( \frac{2 \Lambda}{ m^{ }_\ell} \Bigr) 
  - 1 \biggl]
  + \rmO\Bigl( \frac{1}{\Lambda} \Bigr)
 \;. \la{delta_2}
\ea 
The integral was carried out by substituting $\omega = p x$ whereby
the integrations factorize; integrating over $p$
first; expanding the result in powers of $m_\ell^2/\Lambda^2$; 
and identifying a finite contribution as 
$
 \int_{-1}^{+1} \! {\rm d}x \, 
 \im \bigl[ 
 \bigl( \frac{1}{2(1-x)} + \fr14 \ln\frac{1+x}{1-x} - \frac{i \pi}{4} \bigr)
 \ln
 \bigl( \frac{1}{2(1-x)} + \fr14 \ln\frac{1+x}{1-x} - \frac{i \pi}{4} \bigr)
 \bigr]
 = \pi (\ln 2 - 1)
$.
(Of course the result is easily reproduced 
through numerical integration, 
verifying also that terms argued to be subleading are indeed so.) 

%%%%%%%%%%%%%%%%%%

Let us now see how the result is modified if we ``mistreat'' the soft
momentum domain by carrying out a Taylor expansion in $m_\ell^2$. 
Expanding \eqs\nr{rho0}, \nr{rhos}
while keeping $\lambda \equiv 0^+$ as an infrared regulator
in the denominator, we have
\be
  \hat{\rho}^{ }_0 \stackrel{|\omega| < p}{\simeq} 
   \frac{\pi m_\ell^2 }{4 p \omega (\omega^2 - p^2 - \lambda^2)}
  \;, \quad
  \hat{\rho}^{ }_s \stackrel{|\omega| < p}{\simeq}  
  \frac{\pi m_\ell^2 \omega}{4 p^3 (\omega^2 - p^2 - \lambda^2)}
  \;. \la{rho_cut_ins}
\ee
Then the inserted cut contribution reads (here the correct integration
bounds need to be kept because the infrared domain is not regulated by
$m_\ell^2$)
\ba
 \im \Pi^\rmii{HTL,ins,cut}_\rmii{R} \!\! & \equiv & \!\! 
 - \frac{m_\ell^2}{8\pi} \int_{\frac{M^2}{4\ko}}^{\Lambda} \! {\rm d}p \, p
 \int_{\frac{M^2}{2\ko} - p }^p \! {\rm d}\omega
 \, \Bigl[ \nB{}(\ko) + \fr12 \Bigr]
 \biggl\{ -\frac{1}{p^3} + \frac{\omega (M^2  + \omega^2 - p^2)}
 {2\ko\, p^3 (\omega^2 - p^2 - \lambda^2)} \biggr\} 
 \nn 
 & = & \frac{m_\ell^2}{4\pi} 
 \Bigl[ \nB{}(\ko) + \fr12 \Bigr]
 \Bigl[ \ln\Bigl( \frac{2 \Lambda } {\lambda} \Bigr) - 1  \Bigr]
 + \rmO\Bigl( \frac{1}{\Lambda}, g^4 T^2 \Bigr)
 \;. \la{delta_4}
\ea
For completeness we note that 
the integrand on the first line can also be obtained directly 
from the last term in \eq\nr{fraction}, by taking 
$\omega \equiv \vec{v}\cdot\vec{p}$
as an integration variable instead of $\vec{v}$.

%%%%%%%%%%%%%%%%%%

As expected, \eqs\nr{delta_2} and \nr{delta_4} depend identically
on the ultraviolet cutoff $\Lambda$. The dependence on $\Lambda$ thus
cancels in the difference that plays a role in our actual computation, 
cf.\ \eq\nr{virtual_2}. In the ``fully'' HTL-resummed result of 
\eq\nr{delta_2}, there is no infrared divergence, with the infrared
regime having been regulated by $m_\ell$. In the
``inserted'' HTL result of \eq\nr{delta_4}, there is an infrared 
divergence, but this cancels against a corresponding 
divergence in the inserted pole contribution, as is demonstrated
in \eq\nr{virtual_2}. This cancellation completes the proof of
infrared insensitivity of the observable $\im\Pi^{ }_\rmii{R}$ 
within the HTL setup. 

%%%%%%%%%%%%%%%%%%%%%%%%%%%%%% SECTION %%%%%%%%%%%%%%%%%%%%%%%%%%%%%%%%%
%
\section{Choice of parameters}
\la{app:D}

The physical Higgs mass is set to $m_H = 126$~GeV. 
In order
to convert pole masses and the muon decay constant to 
$\msbar$ scheme parameters at a scale $\bmu = \bmu_0 \equiv m_Z$
we employ 1-loop relations 
specified in ref.~\cite{generic}; 
subsequently, 1-loop renormalization group equations determine
the running of the couplings to a scale
\be
 \bmu_\rmi{ref} \equiv {\rm max}(M,\pi T)
 \;, \la{bmu_ref}
\ee
where they are evaluated for purposes of the present paper. 
Within this approximation the U(1), SU(2) and SU(3) 
gauge couplings $g_1^2,g_2^2,g_3^2$ have explicit solutions
(we have set $\Nc = 3$ and considered 3 families), 
\ba
 g_1^2(\bmu)  =  \frac{48\pi^2}{41 % (22\Nc + 57)
 \ln(\Lambda_1 / \bmu)}
 \;, \quad 
 g_2^2(\bmu)  =  \frac{48\pi^2}{19 % (37 - 6\Nc )
 \ln(\bmu/\Lambda_2)}
 \;, \quad
 g_3^2(\bmu)  =  \frac{24\pi^2}{21 % (11\Nc - 12)
  \ln(\bmu/\Lambda_3)}
 \;, 
\ea
where $\Lambda_1,\Lambda_2,\Lambda_3$ are solved from the boundary 
values at $\bmu = \bmu_0$. The top Yukawa and the Higgs
self-coupling at $\bmu > \bmu_0$ are solved numerically from 
\ba
 \bmu \frac{{\rm d} h_t^2}{{\rm d}\bmu} & = & 
 \frac{h_t^2}{8\pi^2}
 \biggl[
  \fr92 % \Bigl(\Nc + \fr32 \Bigr)
  h_t^2 
   - \frac{17}{12} g_1^2 - \fr94 g_2^2 - 8 % \frac{3(\Nc^2-1)}{\Nc}
 g_3^2     
 \biggr]
 \;, \\ 
 \bmu \frac{{\rm d} \lambda}{{\rm d}\bmu} & = & 
 \frac{1}{8\pi^2}
 \biggl[ 
    \frac{3}{16} \Bigl( g_1^4 + 2 g_1^2 g_2^2 + 3 g_2^4 \Bigr)  
 - \frac{3}{2}\lambda \Bigl( g_1^2 + 3  g_2^2 \Bigr)
 + 12 \lambda^2
 + 6 % 2  \Nc  
  \lambda h_t^2 
 - 3 h_t^4% \Nc 
 \biggr]
 \;. \hspace*{10mm}
\ea
For definiteness let us 
recall that at tree level 
$\lambda \approx g_2^2 m_H^2 / (8 m_W^2) \approx 0.13$.

%%%%%%%%%%%%%%%%%%%%%%%%%%%%%%%%%%%%%%%%%%%%%%%%%%%%%%%%%%%%%%%%%%%%%%%%%%%
%

\end{document}